\documentclass[useAMS,usenatbib]{mnras}
\usepackage[english]{babel}
\usepackage{amsmath}
\usepackage{graphicx}
\usepackage{color}
\usepackage[]{natbib}
\title[Dual and offset AGN in simulations]{Origin and properties of dual and offset active galactic nuclei in a cosmological simulation at z=2}
\author[Steinborn et al.]{Lisa K. Steinborn$^{1}$\thanks{E-mail: steinborn@usm.lmu.de}, Klaus Dolag$^{1,2}$, Julia M. Comerford$^{3}$, Michaela Hirschmann$^{4}$, \and Rhea-Silvia Remus$^{1}$, Adelheid F. Teklu$^{1,5}$
\\
$^{1}$Universit\"ats-Sternwarte M\"unchen, Scheinerstr.1, D-81679 M\"unchen, Germany\\
$^{2}$Max-Planck-Institut f\"ur Astrophysik, Karl-Schwarzschild Strasse 1, D-85740 Garching, Germany\\
$^{3}$Department of Astrophysical and Planetary Sciences, University of Colorado, Boulder, CO 80309, USA\\
$^{4}$UPMC-CNRS, UMR7095, Institut d'Astrophysique de Paris, Boulevard Arago, F-75014 Paris, France\\
$^{5}$Excellence Cluster Universe, Boltzmannstr. 2, D-85748 Garching, Germany}

\begin{document}

\date{Accepted 2016 February 8. Received in original form 2015 October 27}

\pagerange{\pageref{firstpage}--\pageref{lastpage}} \pubyear{2015}

\maketitle

\label{firstpage}

\begin{abstract}
In the last few years, it became possible to observationally resolve
galaxies with two distinct nuclei in their centre.  For separations
smaller than 10kpc, dual and offset active galactic nuclei (AGN) are
distinguished: in dual AGN, both nuclei are active, whereas in offset
AGN only one nucleus is active. To 
study the origin of
such AGN pairs, we employ a cosmological, hydrodynamic simulation with
a large volume of $(182 \mathrm{Mpc})^3$ from the set of Magneticum
Pathfinder Simulations.  The simulation self-consistently produces 35
resolved black hole (BH) pairs at redshift $z=2$, with a comoving
distance smaller than 10kpc.  14 of them are offset AGN and nine are
dual AGN, resulting in a fraction of $(1.2 \pm 0.3)\%$ AGN pairs with
respect to the total number of AGN.  In this paper, we discuss
fundamental differences between the BH and galaxy properties of dual
AGN, offset AGN and inactive BH pairs and investigate their different
triggering mechanisms. 
We find that in dual AGN the BHs have similar masses and the corresponding
BH from the less massive progenitor galaxy always accretes with a higher
Eddington ratio.
In contrast, in offset AGN the active BH is typically more massive than
its non-active counterpart.  Furthermore, dual AGN in general accrete
more gas from the intergalactic medium than offset AGN and non-active
BH pairs. This highlights that merger events, particularly minor
mergers, do not necessarily lead to strong gas inflows and thus, do
not always drive strong nuclear activity.
\end{abstract}

\begin{keywords}
methods: numerical -- galaxies: active -- galaxies: evolution -- galaxies: interactions -- galaxies: nuclei -- quasars: supermassive black holes
\end{keywords}

\section{Introduction}
During the last few years, large-scale hydrodynamic cosmological simulations were established as one of the most powerful tools to make predictions for the evolution of the baryonic structures in the Universe.
Thanks to the increasing power of modern supercomputers, the resolution in these simulations could be increased, helping to resolve the morphological structures of galaxies, even producing disk galaxies
(\citealt{Vogelsberger_nature}, \citealt{Schaye}, \citealt{Remus_2015}, \citealt{Teklu_2015}), which was a long-standing problem in the past.
Since numerical and physical effects appear on scales which were not resolved before, the implemented physical models and numerical schemes have to be refined.
Much progress was made in the field of BH growth.
The original implementation of \citet{Springel_BHs} contains only thermal, radiative AGN feedback,
where the surrounding gaseous medium is heated with a constant efficiency.
However, AGN feedback is known to work in two different regimes, so-called radio- and quasar-mode, or more intuitive, jet- and radiative mode, being dominated by jets and by radiation, respectively.
In general, the two modes represent different phases of super-massive BH (SMBH) growth.
As long as enough gas is driven towards the centre of a galaxy the BH accretes close to the Eddington accretion rate (radiative mode). As soon as the accretion rate decreases due to gas consumption or heating via feedback processes, some AGN form enormous jets, while the radiative power is much less efficient than before (jet-mode).
To account for these different modes of AGN feedback, several authors (e.g. \citealt{Sijacki}, \citealt{Rosas_Guevara}, \citealt{Vogelsberger}, \citealt{Sijacki_2014}, \citealt{Steinborn_2015}) introduced more detailed sub-grid models for cosmological simulations.
These simulations reproduce several observational constraints successfully,
for example the M-$\sigma$ relation and the $M_{\bullet}-M_*$ relation (e.g. \citealt{Magorrian}, \citealt{Silk_Rees_1998}, \citealt{Ferrarese_Merritt_2000}, \citealt{Tremaine}, \citealt{Haering}, \citealt{McConnell}, \citealt{Kormendy}), the AGN luminosity function (e.g. \citealt{Hopkins}) and the BH mass function (e.g. \citealt{Marconi}, \citealt{Shankar04}, \citealt{Shankar09}, \citealt{Shankar13}).

However, while much progress was made in the field of galaxy merger simulations (e.g. \citealt{vanWassenhove_2012}, \citealt{Blecha_2013}, \citealt{Volonteri_2015a}, \citealt{Capelo}) and zoomed-in simulations (e.g. \citealt{Bellovary_seeding}),
in most cosmological simulations the BHs are artificially kept in the centres of the galaxies by pinning them to the gravitational potential minimum (\citealt{Blecha_2015}, \citealt{Springel_BHs}).
The main reason for this treatment is to avoid spurious oscillations of the BHs around the galaxy centres, which appear artificially when the mass of the BH particle is comparable or even smaller than the mass of the surrounding particles.
This issue was addressed by e.g. \citet{Okamoto_2008}, who artificially drag the BHs along the local density gradients instead of pinning them.
\citet{Dubois_2014}, \citet{Tremmel_2015} and \citet{Volonteri_2016} use a similar approach by accounting for a dynamical friction force 
exerted onto
the BHs by the surrounding gas.

Only simulations in which BHs are not pinned back to the potential minimum are able to mimic observations of dislocated BHs (e.g. \citealt{Mueller_Sanchez_2015}, \citealt{Allen_2015}, \citealt{Comerford_2015}, \citealt{Comerford_2012}, \citealt{McGurk_2011}, \citealt{Barth_2008}),
which are thought to originate from galaxy merger events.
When both progenitor galaxies host an SMBH, a close pair of SMBHs can form.
At the stage at which these BHs are separated by a few kpc, they can be observed as dual or offset AGN, which are also called proto super-massive binary BHs in the literature \citep{Hudson}.
In dual AGN, both BHs are observed as AGN, whereas in offset AGN only one BH is active.
Note that the expression `offset AGN' is also used in literature for single AGN which are offset from the galaxy centre.
In these cases, the offsets are thought to be caused by recoiling BHs after BH mergers (e.g. \citealt{Sijacki_2011}, \citealt{Volonteri_Madau}, \citealt{Blecha_2015}).
In contrast to BH pairs, recoiling BHs are not self-consistently produced by cosmological simulations.
Following \citet{Comerford_2015}, dual and offset AGN are BH pairs with a spatial separation of less than 10kpc.
However, other maximum separations up to 100kpc exist in the literature \citep{Koss_2012}.

In the local Universe, offset AGN may be quite common (\citealt{Comerford_Greene}, \citealt{Comerford_2009b}, \citealt{Comerford_2013}), while dual AGN might be more rare (e.g. \citealt{Rosario_2011}, \citealt{Fu_Myers}).
In addition, they are still difficult to observe due to the small spatial separations.
Although first evidence for the existence of dual AGN was already found by \citet{Owen}, who observed two distinct radio jets in the radio source 3C 75 in Abell 400,
only a few dual AGN have been confirmed so far.
To confirm dual AGN observationally, both nuclei need to be spatially resolved (\citealt{Komossa}, \citealt{Hudson}, \citealt{Bianchi}, \citealt{Koss_2011}, \citealt{Koss_2012}, \citealt{Mazzarella}, \citealt{Shields}).
One possibility to find dual AGN is to search for double-peaked narrow AGN emission lines.
However, a double-peak alone can also be produced by the kinematics of the narrow line region of a single AGN \citep{Mueller_Sanchez_2015} and hence this method can only be used to select dual AGN candidates (e.g. \citealt{Comerford_2009}, \citealt{Comerford_2011}, \citealt{Barrows_2013}).
Using this method, \citet{Fu_Zhang}, \citet{Liu}, \citet{Comerford_2015} and \citet{Mueller_Sanchez_2015} recently found seven dual AGN systems, in which the existence of dual AGN
was
subsequently confirmed by spatially resolving two distinct nuclei with separations less than 10kpc.

These detections indicate that galaxy mergers might trigger AGN activity, but since observations only capture ``one moment in time", it is still unclear whether the AGN are actually triggered by the gas inflow due to merger events or whether they have already been luminous before the merger event, i.e., whether their nuclear activity is driven by internal processes or due to their location in the large-scale cosmic web.
Related to that, it is also a matter of vigorous debate why in some BH pairs both BHs are active and in others only one of them or even none is active.

To investigate the origin of the differences between dual AGN, offset AGN, and inactive BH pairs and to explore the underlying driving mechanisms for AGN activity during galaxy mergers, we employ a large-scale cosmological simulation at redshift $z=2$ with both a large volume of $(182\mathrm{Mpc})^3$ and sufficiently high resolution to properly resolve the morphology of galaxies.
In this simulation, we do not artificially keep BHs in the galaxy centre, providing a first attempt to study BH pairs in a fully cosmological context, as suggested by \citet{Volonteri_IAU_2015}.

The paper is structured as follows. In Section \ref{simulation}, we briefly introduce our simulation and the BH model. 
We present our sample of BH pairs in Section \ref{sample}
and discuss our results in Section \ref{results}.
In Section \ref{comparison}, we compare our results to other theoretical studies and finally, we summarize our main results in Section \ref{conclusion}.

\section{The simulation}
\label{simulation}
\subsection{Set-up}
We use a simulation taken from the Magneticum Pathfinder Simulation set\footnote{www.magneticum.org} (Dolag et al. in prep.),
which is an ensemble of cosmological simulations including full hydrodynamics based on the TreePM-SPH code P-GADGET3 \citep{Springel}.
We assume a standard $\Lambda$CDM cosmology, where the Hubble parameter is $h=0.704$ and the density parameters for matter, dark energy and baryons are $\Omega_\mathrm{m}=0.272$, $\Omega_\mathrm{\Lambda}=0.728$ and $\Omega_\mathrm{b}=0.0451$, respectively (WMAP7, \citealt{Komatsu}).
The Magneticum Simulations contain sub-resolution models for star formation \citep{Springel_Hernquist}, isotropic thermal conduction
\citep{Dolag04} with an efficiency of $\kappa=1/20$ of the classical
{\it Spitzer} value \citep{Arth}, stellar evolution, metal enrichment, and supernova
feedback (\citealt{Tornatore03}, \citealt{Tornatore}) as well as a cooling
function which depends on the individual metal species following
\citet{Wiersma}.
Furthermore, we use an updated viscosity treatment (\citealt{Dolag05}, \citealt{Beck_2015}) and Wendland kernels from \citet{Dehnen}, see \citet{Donnert} and \citet{Beck_2015}.

We use the same simulation set-up as described in \citet{Steinborn_2015}, 
but a higher resolution, with particle masses $M_{\textrm{dm}}=3.6 \cdot 10^7 M_{\odot}/h$ 
for dark matter, $M_{\textrm{gas}}=7.3 \cdot 10^6 M_{\odot}/h$ as initial 
mass of a gas particle and $M_{\textrm{stars}}\approx1.8 \cdot 10^6 M_{\odot}/h$ 
as typical mass for a star particle. 
The gravitational softening length for dark matter particles was kept fixed
at $4.2 \mathrm{kpc}/h$ comoving Plummer-equivalent and was switched to
a physical softening length of $1.4 \mathrm{kpc}/h$ at $z=2$ while the
softening length for gas and star particles was fixed to 
$1.4 \mathrm{kpc}/h$ and $0.7 \mathrm{kpc}/h$ comoving Plummer-equivalent,
respectively. The black hole sink particles are treated with the same softening
as star particles and we did not apply any restriction to the SPH smoothing
length of the gas particles. This allows
us to resolve BH pairs roughly down to 2kpc, which is twice as large
as the softening length of the stars (see \citealt{Teklu_2015} and
Remus et al. in prep. 2016 for a detailed study of galaxy properties
at this resolution level). We remark that this is a very
conservative assumption for the resolution and hence, although we
cannot give a prediction for the number of BH pairs with distances below 2kpc,
the simulation can still contain BH pairs with smaller spatial separations.
To get a representative sample of BH pairs, we performed a large
simulation with a volume of $(182\mathrm{Mpc})^3$ and an initial
particle number of 7.2 billion particles down to redshift $z=2$, where
we find many BH pairs due to the rather high merger rate at that time.

\subsection{Model for BH growth}
The implementation of BH physics is based on \citet{Springel_BHs}, in
which BHs are treated as collision-less sink particles.
We evaluate on the fly friends-of-friends (FoF) groups of star particles
with a seven times smaller than usual linking-length to identify stellar 
objects (e.g. galaxies) which do not yet contain black holes.
If the mass of such stellar objects is larger than $4\cdot
10^{9} M_{\odot}/h$, the star particle with the highest binding energy 
is replaced by a black hole sink particle. For the initial true mass of 
the BH particles we choose a BH mass which is clearly below the mass
expected from the observed relation between the BH mass and the stellar mass
(e.g. \citealt{McConnell}), i.e. $M_{\bullet}=5 \cdot 10^5
M_{\odot}/h$. In that way, we avoid an overheating of the
surrounding gas, since the effect of the AGN feedback is less
abrupt. For the dynamical mass of the BH sink particle we choose the mass
of a single dark matter particle. The dynamical mass of the sink
particle is only used as mass when computing the gravitational forces and 
it only starts to increase
when the true mass of the BH particle
starts to grow and exceeds the dynamical mass.

In contrast to the original implementation from
\citet{Springel_BHs}, we use a more detailed description of AGN
feedback and BH accretion according to \citet{Steinborn_2015}. In
this model, we account for both radiation and mechanical outflows,
which are implemented as thermal feedback since the scales on
which mechanical feedback would act are not resolved.  The
radiative efficiency $\epsilon_{\mathrm{r}}$ as well as the efficiency
of mechanical outflows $\epsilon_{\mathrm{o}}$ is variable and depends
on the accretion rate onto the BH $\dot{M}_{\bullet}$ and on the BH
mass $M_{\bullet}$, as deduced from observations
(\citealt{Davis}, \citealt{Chelouche}).  We fix the coupling factor of
the radiation to the surrounding medium to
$\epsilon_{\mathrm{f}}=0.03$.  This value is chosen such that the
simulation reproduces the normalization of the observed
$M_{\bullet}$-$M_*$ relation from \citet{McConnell}. Since this factor
depends on the resolution, it is clearly lower than the value
$\epsilon_{\mathrm{f}}=0.2$ from \citet{Steinborn_2015}, where a simulation with lower resolution was studied.

The accretion rate is limited to the Eddington accretion rate $\dot{M}_{\mathrm{Edd}}$. Furthermore we distinguish between hot ($T>5\cdot 10^5$K) and cold ($T<5\cdot 10^5$K) gas accretion:
\begin{equation}
    \dot{M_\bullet} = \mathrm{min}(\dot{M}_\mathrm{B, hot} +
    \dot{M}_\mathrm{B, cold}, \dot{M}_{\mathrm{Edd}}) ,
\end{equation}
where both $\dot{M}_\mathrm{B, cold}$ and $\dot{M}_\mathrm{B, hot}$ are computed independently from each other using the Bondi accretion rate (\citealt{Hoyle},
\citealt{Bondi}, \citealt{Bondi_Hoyle}), which is given by
\begin{equation}
\dot{M}_\mathrm{B} = \frac{4 \pi \alpha G^2 M_\bullet^2 \rho_{\infty}}{(v^2 + c_{\mathrm{s}}^2)^{3/2}},
\label{accretion_rate}
\end{equation}
where $\rho$ is the density, $c_s$ is the sound speed of the accreted
gas and $v$ is the velocity of the gas relative to that of the black hole.
To account for turbulence (following \citealt{Gaspari}), we multiply the cold gas accretion rate by the boost factor $\alpha=100$ and the hot gas accretion rate by $\alpha=10$.
For further details regarding the sub-grid model for BH accretion and AGN feedback, see \citet{Steinborn_2015}.

Furthermore, we do not perform any pinning of the BHs to the deepest
potential in their surrounding, but let them evolve
self-consistently. This prevents the BH sink
particles from merging too early and is only possible due to an improved numerical handling
of the BH sink particles, which effectively avoids artificial drifting
of such BH sink particles due to numerical inaccuracies as observed in
the past. Thereby, the BH sink particles evolve self-consistently
during a merger event. In particular, seeding on the star particle 
with the highest binding energy (compared to choosing the dark matter particle
with the largest density as originally done in \citealt{Springel_BHs}) already
leads to a very precisely centered BH sink particle from the beginning.
Additionally, by setting the dynamical mass of the BH sink particle to
the mass of a dark matter particle, by carefully choosing the softening,
and by seeding only in galaxies which are resolved by the order of
thousands of star particles, we guarantee that the code naturally captures dynamical friction and no additional friction term
has to be used at this resolution.

Finally, we modified the conditions under which two BH sink particles will merge.
Close pairs of BH sink particles will not be merged as long as they fulfill one of the following conditions:
(i) the relative velocity of the BHs to each other is larger than $0.5 c_{\mathrm{s}}$, where $c_{\mathrm{s}}$ is the sound-speed of the surrounding gas;
(ii) the distance is larger than five times the softening length and none of the two BHs is gravitationally bound to the other one.
For too large separations between the
BHs the second argument prevents a too early merging of the two BH sink particles.
This approach still ignores any additional
time the BHs need for the final merging, assuming that this timescale is still smaller than the dynamical time-steps of the cosmological simulation.
Nevertheless, due to our novel approach in handling the BH sink particles, the final merging of the BHs in the simulation happens only in the very late state of the merger event, allowing us to
study, for the first time, close BH pairs in cosmological simulations.

\subsection{AGN luminosities}
\label{AGN_luminosities}
\begin{figure}
  \includegraphics[trim = 11mm 76mm 17mm 86mm, clip,width=0.5\textwidth]{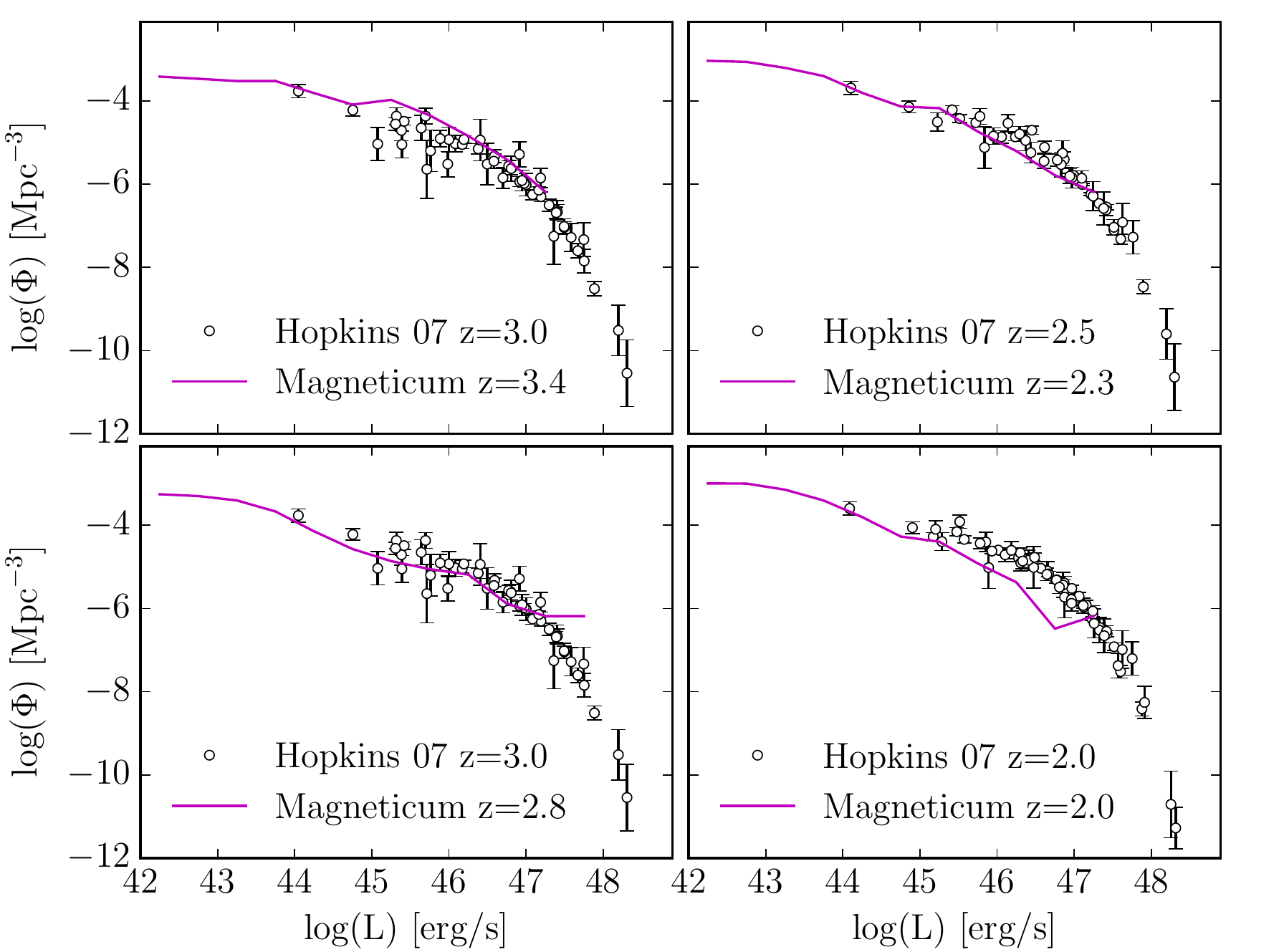}
  \caption{Bolometric AGN luminosity function of our simulation (solid magenta lines) at different redshifts in comparison to the observations (black circles with error bars) from \citet{Hopkins}.
          }
\label{LF}
\end{figure}
The bolometric AGN luminosities of our simulated BHs are calculated according to
\begin{equation}
L=\epsilon_{\mathrm{r}} \dot{M}_{\bullet} c^2.
\end{equation}

Fig. 1 shows the bolometric AGN luminosity function at different
redshifts. Our simulation (solid magenta lines) is successful in
reproducing the observations from \citet{Hopkins} at roughly the same redshifts (black circles with
error bars), particularly between $z = 3.4$ and $z = 2.3$.  At 
$z=2$
the simulation slightly underestimates the number density
of AGN, in particular at the very bright end of the luminosity
function.  
However, this is most likely
due to the inefficient
size of the simulation volume. Our simulation at $z=2$ contains only
19 AGN with $L_\mathrm{bol}>10^{46}$erg/s, so a much larger volume would be
needed to properly capture the bright end of the AGN luminosity
function (see also \citealt{Hirschmann}).  Nevertheless, the
simulation volume is large enough to produce a statistically realistic
population of AGN and therefore provides a good base to study dual and
offset AGN for the first time in cosmological simulations.

\section{Sample of BH pairs}
\label{sample}
Our simulation contains a total number of 14,903 BHs at $z=2$ which
are more massive than $10^7 M_{\odot}$. Seeding galaxies
with BHs marks an abrupt change of the galaxy formation
physics. As described in Section \ref{simulation}, the seed mass of our BHs
is chosen to be significantly smaller than expected for the stellar mass
of the galaxy. The BH will then grow and evolve into a self-regulated
state and thereby release the associated feedback energy, compensating partially
this missing feedback in the early evolution of these galaxies. BHs which did not yet evolve into this self-regulated state are referred to as 
`unresolved BHs'. The chosen mass threshold of $10^7 M_{\odot}$ corresponds roughly to the cut-off
in the BH mass function for the high-resolution simulation shown in \citet{Hirschmann}.
\footnote{We choose a relatively low BH mass threshold, since
this cut-off and thus the transition from resolved to unresolved
BHs, is very smooth for the high-resolution simulation.}
For lower BH masses, it is possible that the system did not yet have time to evolve into a self-regulated state and therefore, the luminosity of the BH
is difficult to interpret.

We identify a sample of 34 BH pairs with a comoving distance smaller
than 10kpc at $z=2$, where at least one BH is resolved in mass.
Throughout this paper, we define a BH as an AGN if it has a bolometric
luminosity larger than $10^{43}$ erg/s. We distinguish between four
different classes of BH pairs:
\begin{itemize}
\item dual AGN,
\item offset AGN (both BHs are more massive than $10^7 M_{\odot}$),
\item unresolved offset AGN (while the AGN is more massive than $10^7
  M_{\odot}$, the second BH is below this resolution limit), and
\item dual BHs without AGN.
\end{itemize}
We distinguish between resolved and unresolved offset AGN,
because it is not clear to which class unresolved offset AGN would
belong if both BH masses were resolved.  Since the unresolved,
non-active BH should actually be more massive, and since the
luminosity depends on the BH mass, BH pairs in this class would be
either classified as dual AGN or offset AGN, if the masses of both BHs
were resolved. The reason for the too low BH mass is not only
the progenitor galaxy itself, but also the more massive counterpart,
which suppresses BH accretion, as we will describe in Section
\ref{triggering_vs_suppressing} in more detail.  Thus, although the
BHs and their environment might not be totally realistic, it is
still an interesting question why the smaller BH could not grow as
much as the larger one. For that reason and to have a
sample as complete as possible, we also consider unresolved offset
AGN in this analysis. We remark that there are also two dual AGN
pairs and several BH pairs without AGN where one BH is below the
mass resolution limit. We do not treat them as a separate class for
the following reasons: (i) if a BH below the mass resolution limit
is active and thus grows properly, we do not treat it as unresolved,
because the AGN feedback is a self-regulated process, which depends
mainly on the gas properties and not on the BH mass;
(ii) if a BH in a
pair without AGN is not resolved, but its counterpart is massive
enough to be resolved and is also not active, we expect that, since
the environment of the two BHs is roughly the same, the unresolved
BH would also be inactive if it was more massive.

Our chosen luminosity threshold is clearly arbitrary and leads to some overlap between the properties of the different classes of BH pairs.
We refer to Section \ref{comparison}, in particular Fig. \ref{Lthresh}, where we show the contribution of the four different classes for different luminosity thresholds.
However, we want to emphasize that a different definition does not change our results qualitatively.

That luminosity threshold gives us 1864 AGN\footnote{Dual AGN pairs are
  counted as one AGN to be consistent with observations.} and our sample of 34 BH pairs then splits up into:
\begin{itemize}
\item{9 dual AGN pairs ($\sim$0.5\% of all AGN),}
\item{6 offset AGN ($\sim$0.3\% of all AGN),}
\item{8 unresolved offset AGN ($\sim$0.4\% of all AGN) and}
\item{11 dual BHs without AGN.}
\end{itemize}
The fraction of dual and offset AGN with respect to the total amount of AGN 
then sums up to $\sim 1.2$\%.
The fraction of dual AGN varies between $\sim 0.48$\% and $\sim 0.91$\%, accounting
for the fact that the unresolved offset AGN might actually be dual
AGN.

\begin{figure}
  \includegraphics[trim = 10mm 60mm 0mm 70mm, clip,width=0.5\textwidth]{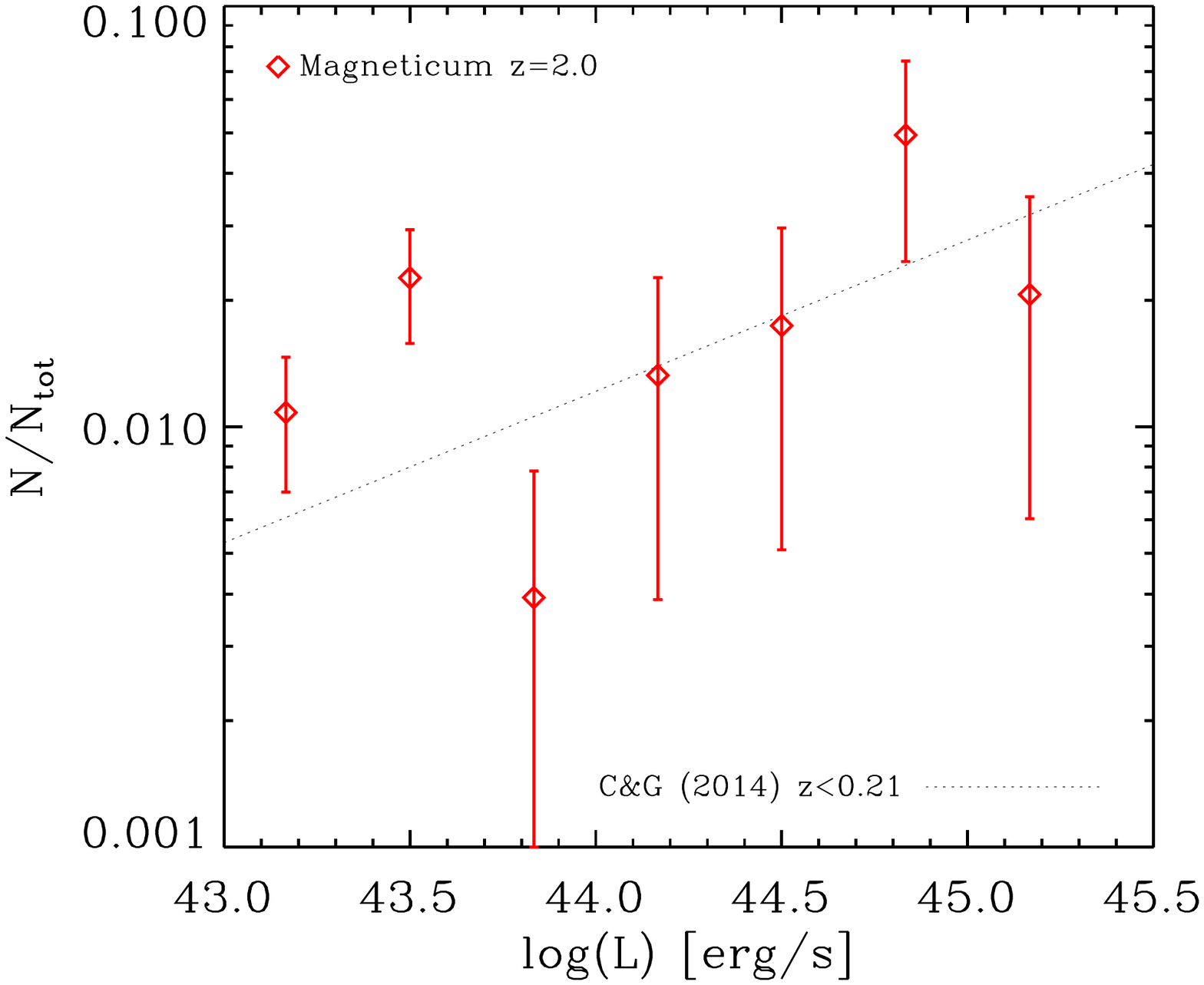}
  \caption{The red diamonds show the fraction of AGN in pairs with
    respect to the total number of AGN in bins of the bolometric
    luminosity, where N is the number of offset and dual AGN in bins
    of the bolometric luminosity and $N_{\mathrm{tot}}$ is the total
    number of AGN in the same luminosity range. The error bars show
    the corresponding $\sqrt{N}/N_{\mathrm{tot}}$ error.
    We see a trend of an increasing fraction of AGN pairs with luminosity. Since Comerford \&
    Greene (C\&G, 2014) found a similar trend in their observations 
    at low redshifts, we also show for orientation
    their best fit for observed offset AGN candidates at $z<0.21$ as black dotted line, although a direct comparison at the different redshifts is not possible.
    }
\label{N}
\end{figure}

Fig. \ref{N} shows the fraction of dual and offset AGN with respect to
the total amount of AGN for a given luminosity bin. The simulation
predictions for $z=2$ are shown as red diamonds, where the red error
bars illustrate the corresponding $\sqrt{N}/N_{\mathrm{tot}}$
error. 
Despite the rather large scatter due to our low number of AGN pairs we
see a trend of an increasing fraction of AGN pairs with luminosity.
Since our simulation ran only down to $z=2.0$, we cannot
directly compare it with observations, which are only available for
very low redshifts. Nevertheless, observations at low redshifts ($z<0.21$, \citealt{Comerford_Greene})
find the same trend as we see in our simulation at $z=2$, in the sense that the amount of
candidate offset AGN increases with AGN luminosity (see black dotted line in Fig. \ref{N}). This could indicate that this trend with luminosity is already in place at $z=2$.
Furthermore, these findings might be directly connected to other observational results showing
that the fraction of AGN triggered by galaxy mergers seems to increase with AGN luminosity
(e.g. \citealt{Treister}).

\begin{table*}
\centering
\begin{tabular} {|l|l|l|l|l|l|l|l|}
\hline
ID & $d$ [kpc] & log($L_1$) [erg/s] & log($L_2$) [erg/s] 
& $M_{\bullet 1} [M_{\odot}]$ & $M_{\bullet 2} [M_{\odot}]$ & $f_{\mathrm{Edd} 1}$ & $f_{\mathrm{Edd} 2}$\\
\hline
1 & 9.68 & 43.58 & 43.57 & $5.94\cdot 10^7$ & $4.42\cdot 10^7$ & $5.14\cdot 10^{-3}$ & $6.73\cdot 10^{-3}$\\
2 & 8.78 & 43.20 & 43.06 & $2.90\cdot 10^7$ & $4.16\cdot 10^7$ & $4.38\cdot 10^{-3}$ & $2.20\cdot 10^{-3}$\\
3 & 2.40 & 43.20 & 43.17 & $2.88\cdot 10^7$ & $1.84\cdot 10^7$ & $4.40\cdot 10^{-3}$ & $6.41\cdot 10^{-2}$\\
4 & 2.67 & 44.77 & 44.76 & $4.49\cdot 10^7$ & $3.62\cdot 10^7$ & $1.04\cdot 10^{-1}$ & $1.27\cdot 10^{-1}$\\
5 & 8.91 & 45.32 & 43.49 & $1.39\cdot 10^8$ & $2.04\cdot 10^7$ & $1.21\cdot 10^{-1}$ & $1.20\cdot 10^{-2}$\\
6 & 3.50 & 43.42 & 43.33 & $1.52\cdot 10^8$ & $1.19\cdot 10^8$ & $1.37\cdot 10^{-3}$ & $1.43\cdot 10^{-3}$\\
7 & 8.25 & 44.53 & 43.35 & $3.97\cdot 10^8$ & $1.59\cdot 10^8$ & $6.83\cdot 10^{-3}$ & $1.12\cdot 10^{-3}$\\
\hline
8 & 7.11 & 44.46 & 44.06 & $3.79\cdot 10^7$ & $5.10\cdot 10^6$ & $6.11\cdot 10^{-2}$ & $1.81\cdot 10^{-1}$\\
9 & 7.17 & 44.72 & 43.20 & $2.35\cdot 10^7$ & $8.98\cdot 10^6$ & $1.79\cdot 10^{-1}$ & $1.40\cdot 10^{-2}$\\
\hline
\end{tabular}
\caption{Properties of the dual AGN pairs in our simulation at $z=2.0$.
	The indices 1 and 2 always correspond to the more and less luminous BH, respectively.
	The two dual AGN below the horizontal line contain one BH with $M_{\bullet}<10^7 M_{\odot}$.
	}
\label{dual_AGN_table}
\end{table*}

\begin{table*}
\centering
\begin{tabular} {|l|l|l|l|l|l|l|l|}
\hline
ID & $d$ [kpc] & log($L_1$) [erg/s] & log($L_2$) [erg/s] 
& $M_{\bullet 1} [M_{\odot}]$ & $M_{\bullet 2} [M_{\odot}]$ & $f_{\mathrm{Edd} 1}$ & $f_{\mathrm{Edd} 2}$\\
\hline
10 & 5.05 & 43.59 & 41.82 & $1.57\cdot 10^8$ & $1.62\cdot 10^7$ & $1.97\cdot 10^{-3}$ & $3.26\cdot 10^{-4}$\\
11 & 9.23 & 43.26 & 42.65 & $6.94\cdot 10^7$ & $4.29\cdot 10^7$ & $2.09\cdot 10^{-3}$ & $8.37\cdot 10^{-4}$\\
12 & 5.91 & 43.01 & 41.80 & $2.72\cdot 10^7$ & $2.52\cdot 10^7$ & $3.00\cdot 10^{-3}$ & $1.99\cdot 10^{-4}$\\
13 & 6.85 & 44.20 & 42.88 & $5.14\cdot 10^7$ & $4.38\cdot 10^7$ & $2.44\cdot 10^{-2}$ & $1.37\cdot 10^{-3}$\\
14 & 4.83 & 43.49 & 40.16 & $5.71\cdot 10^8$ & $1.08\cdot 10^8$ & $4.30\cdot 10^{-4}$ & $1.73\cdot 10^{-6}$\\
15 & 7.91 & 43.59 & 42.37 & $1.49\cdot 10^8$ & $3.13\cdot 10^7$ & $2.08\cdot 10^{-3}$ & $5.88\cdot 10^{-4}$\\
\hline
16 & 8.88 & 44.68 & 37.23 & $3.52\cdot 10^8$ & $1.24\cdot 10^6$ & $1.07\cdot 10^{-2}$ & $1.10\cdot 10^{-7}$\\
17 & 1.40 & 43.44 & 41.79 & $1.04\cdot 10^7$ & $6.19\cdot 10^6$ & $2.10\cdot 10^{-2}$ & $7.89\cdot 10^{-4}$\\
18 & 7.76 & 43.77 & 41.51 & $4.57\cdot 10^7$ & $2.51\cdot 10^6$ & $1.02\cdot 10^{-2}$ & $1.03\cdot 10^{-3}$\\
19 & 7.41 & 43.40 & 40.98 & $1.35\cdot 10^7$ & $2.41\cdot 10^6$ & $1.48\cdot 10^{-2}$ & $3.12\cdot 10^{-4}$\\
20 & 9.16 & 43.39 & 41.82 & $4.59\cdot 10^7$ & $3.21\cdot 10^6$ & $4.23\cdot 10^{-3}$ & $1.65\cdot 10^{-3}$\\
21 & 5.94 & 44.03 & 39.79 & $1.04\cdot 10^8$ & $2.50\cdot 10^6$ & $8.20\cdot 10^{-3}$ & $1.97\cdot 10^{-5}$\\
22 & 5.59 & 43.14 & 39.34 & $2.92\cdot 10^7$ & $1.22\cdot 10^6$ & $3.74\cdot 10^{-3}$ & $1.42\cdot 10^{-5}$\\
23 & 6.20 & 45.21 & 39.66 & $1.36\cdot 10^7$ & $1.23\cdot 10^6$ & $9.45\cdot 10^{-1}$ & $2.98\cdot 10^{-5}$\\
\hline
\end{tabular}
\caption{Same as table \ref{dual_AGN_table} but for resolved (upper 6 lines) and unresolved (lower 8 lines) offset AGN.
	}
\label{offset_AGN_table}
\end{table*}

In Tables \ref{dual_AGN_table} and \ref{offset_AGN_table} we summarize the most important properties of the simulated dual and offset AGN pairs, where $d$ is the comoving distance of the BHs to each other and the indices 1 and 2 correspond to the more and less luminous AGN.
We see no dependence of the luminosities on the distance between the BHs.
We calculated the luminosities $L_1$ and $L_2$ of the two BHs as described in \citet{Steinborn_2015}.
The Eddington ratios $f_{\mathrm{Edd} 1}$ and $f_{\mathrm{Edd} 2}$ are defined as $f_{\mathrm{Edd}}:=L/L_{\mathrm{Edd}}$.
\footnote{Note that in our model $L/L_{\mathrm{Edd}}$ is not the same as $\dot{M}_{\bullet}/M_{\mathrm{Edd}}$ \citep{Steinborn_2015}, since the total accretion rate splits up into luminosity and mechanical outflows: $\dot{M}_{\bullet}/M_{\mathrm{Edd}} = L/L_{\mathrm{Edd}} + P_o/L_{\mathrm{Edd}}$, where $L=\epsilon_r\dot{M}_{\bullet}c^2$ is the luminosity and $P_o=\epsilon_o\dot{M}_{\bullet}c^2$ is the power of mechanical outflows.}

\begin{figure}
  \includegraphics[trim = 10mm 60mm 0mm 65mm, clip,width=0.5\textwidth]{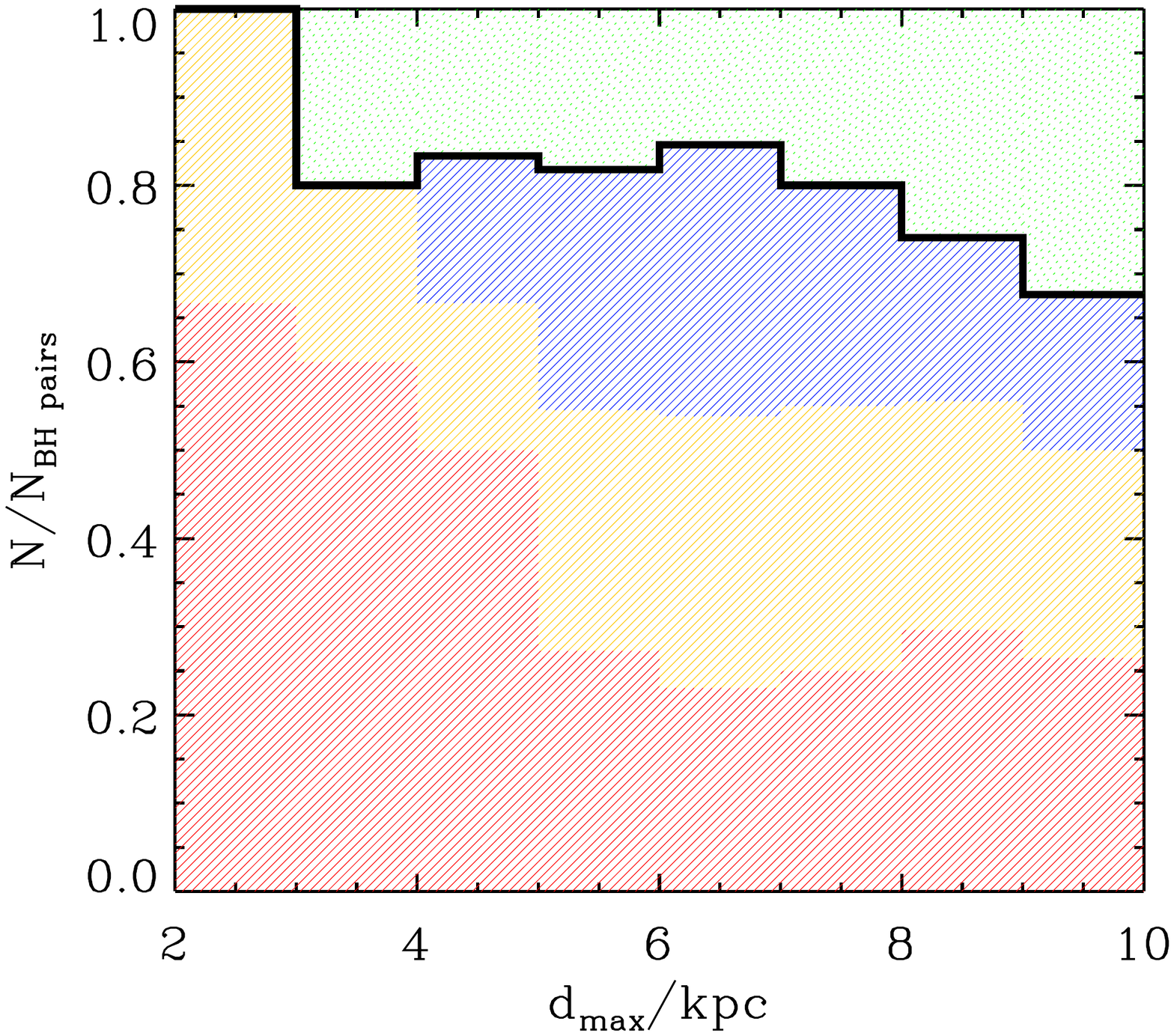}
  \caption{The black line shows the cumulative fraction of dual and offset AGN in bins of the maximum separation $d_{\mathrm{max}}$ between the two BHs. The coloured areas indicate the contribution from dual AGN (red), offset AGN (blue), unresolved offset AGN (yellow) and BH pairs without AGN (green).
          }
\label{d}
\end{figure}

Fig. \ref{d} shows the cumulative fraction of dual and offset AGN for a given maximum separation $d_{\mathrm{max}}$ between the two BHs.
The coloured areas indicate the contribution from the four different classes of BH pairs: dual AGN (red), offset AGN (blue), unresolved offset AGN (yellow) and BH pairs without AGN (green).
Although our sample size is rather small, it is clearly visible that dual AGN dominate at small separations, whereas the fraction of BH pairs without AGN is the largest when allowing separations up to 10kpc.
Furthermore, all resolved offset AGN have separations larger than 4kpc.
This is also the case for the unresolved offset AGN, except for one example with an extremely low separation.
Overall, our simulation predicts the same trend at $z=2$ like the observations from \citet{Comerford_2015} at lower redshifts, which indicate that the fraction of AGN increases with decreasing BH distances.
Such a trend was also observed much more accurately for larger separations between 10kpc and 100kpc (\citealt{Ellison_2011}, \citealt{Koss_2012}).

\begin{figure*}
  \includegraphics[trim = 0mm 0mm 0mm 0mm, clip,width=0.87\textwidth]{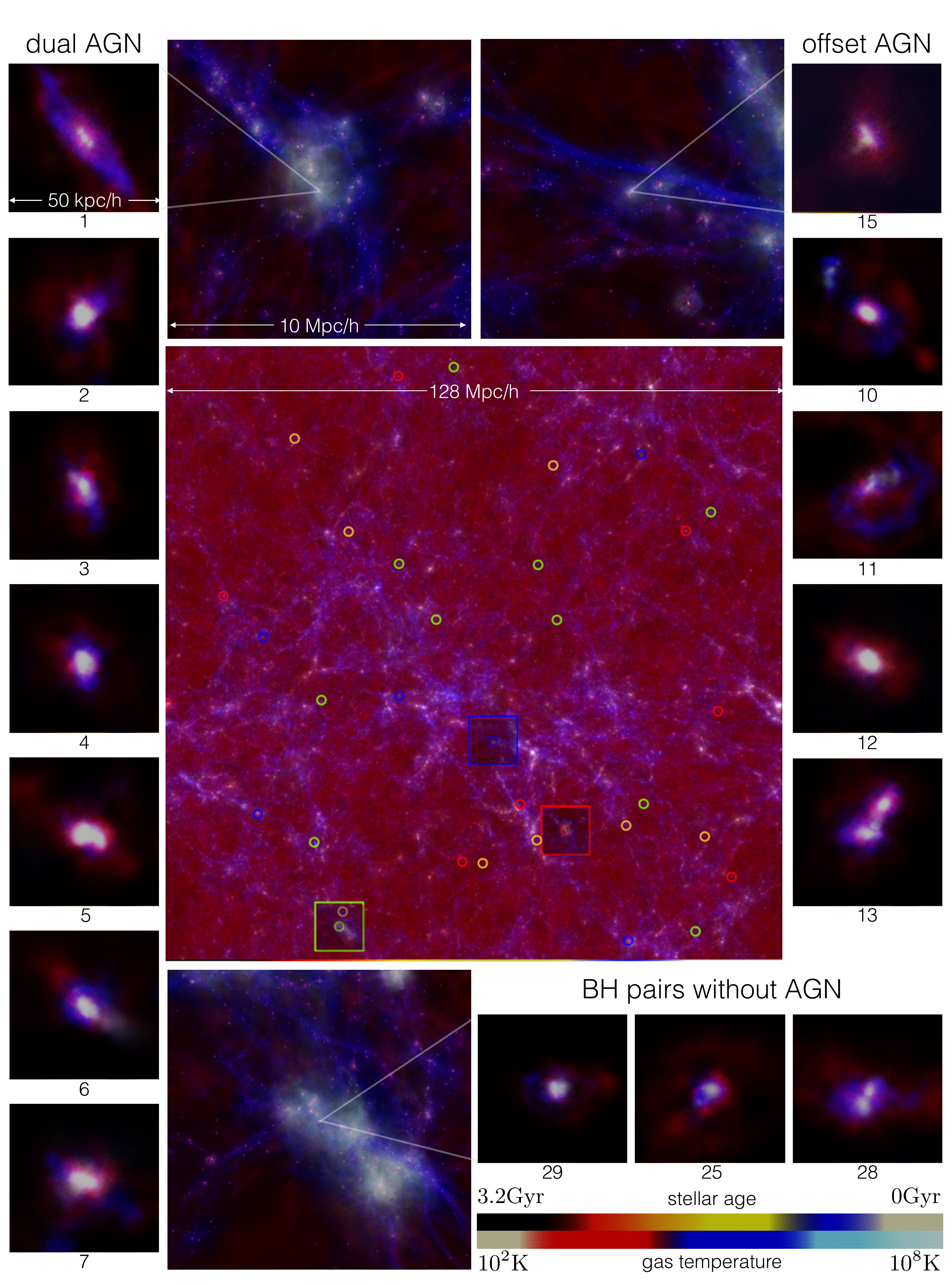}
  \caption{The large panel in the middle shows a visualization of our cosmological simulation.
    The red, blue and green circles mark the positions of all dual AGN pairs, offset AGN and BH pairs without AGN, respectively.
    Exemplarily, we show the large-scale environment, i.e. a box of 10 Mpc/h length around the host galaxy of one dual AGN pair, one offset AGN and one BH pair without AGN.
    The positions of these boxes are also marked in the large picture.
    We remark that the box is so large that structures are often not visible because they are overlaid by something else.
    Furthermore, we show a few examples of the host galaxies of the dual AGN (left images), offset AGN (right images) and BH pairs without AGN (images in the middle bottom), where we always show a box with a length of {50} kpc/h.
    The IDs of the BH pairs are the same as in Table \ref{dual_AGN_table} and \ref{offset_AGN_table} and the numbering continues to BH pairs without AGN (IDs 24-34).
    The colour bars are the same for all pictures,
    where the upper colour bar represents the age of the stars (from old to young in logarithmic scale of the cosmic age $a$, converted to the stellar age) and the lower one the gas temperature (from cold to hot in logarithmic scale).
	  }
\label{collage}
\end{figure*}
In Fig. \ref{collage}, we show visualizations\footnote{performed with the free software Splotch, http://www.mpa-garching.mpg.de/$\sim$kdolag/Splotch from \citet{Dolag_splotch}} of a few examples of the three different classes of AGN/BH pairs.
The images illustrate the baryonic component only. Gas and stars are colour-coded by the gas temperature and the stellar age, respectively (as indicated by the colour bars).
In the middle we show the total box. To check whether the different BH pairs are located in different large scale environments, we highlight the positions of all BH pairs as coloured circles, where the colours red, blue, yellow and green represent dual AGN pairs, offset AGN, unresolved offset AGN and BH pairs without AGN, respectively.
At first sight, there seems to be no obvious difference between the environment of the different types. However, this could be a consequence of overlaid structures due to the large volume.
Hence, for three examples, we additionally show a smaller box of $10 \mathrm{Mpc}/h$ around the BH pair.
We show larger illustrations of these regions in the three medium sized images at the middle top and middle bottom.  
The panel at the bottom demonstrates, for example, that non-active BHs are not necessarily located in a gas poor environment.
Overall, we can suspect that, in contrast to offset AGN and inactive BH pairs, dual AGN (left large image at the top) have a higher probability to be located in the centre of large-scale filaments, which can provide the gas supply for the galaxy.
In Section \ref{trace} we will address this issue in more detail.

\section{Results}
\label{results}
In this section, we investigate different properties of the BH pairs to understand the underlying mechanisms leading to the differences in their AGN activity.
\subsection{BH and stellar masses}
\label{properties}
\begin{figure}
  \includegraphics[trim = 0mm 60mm 0mm 60mm, clip,width=0.5\textwidth]{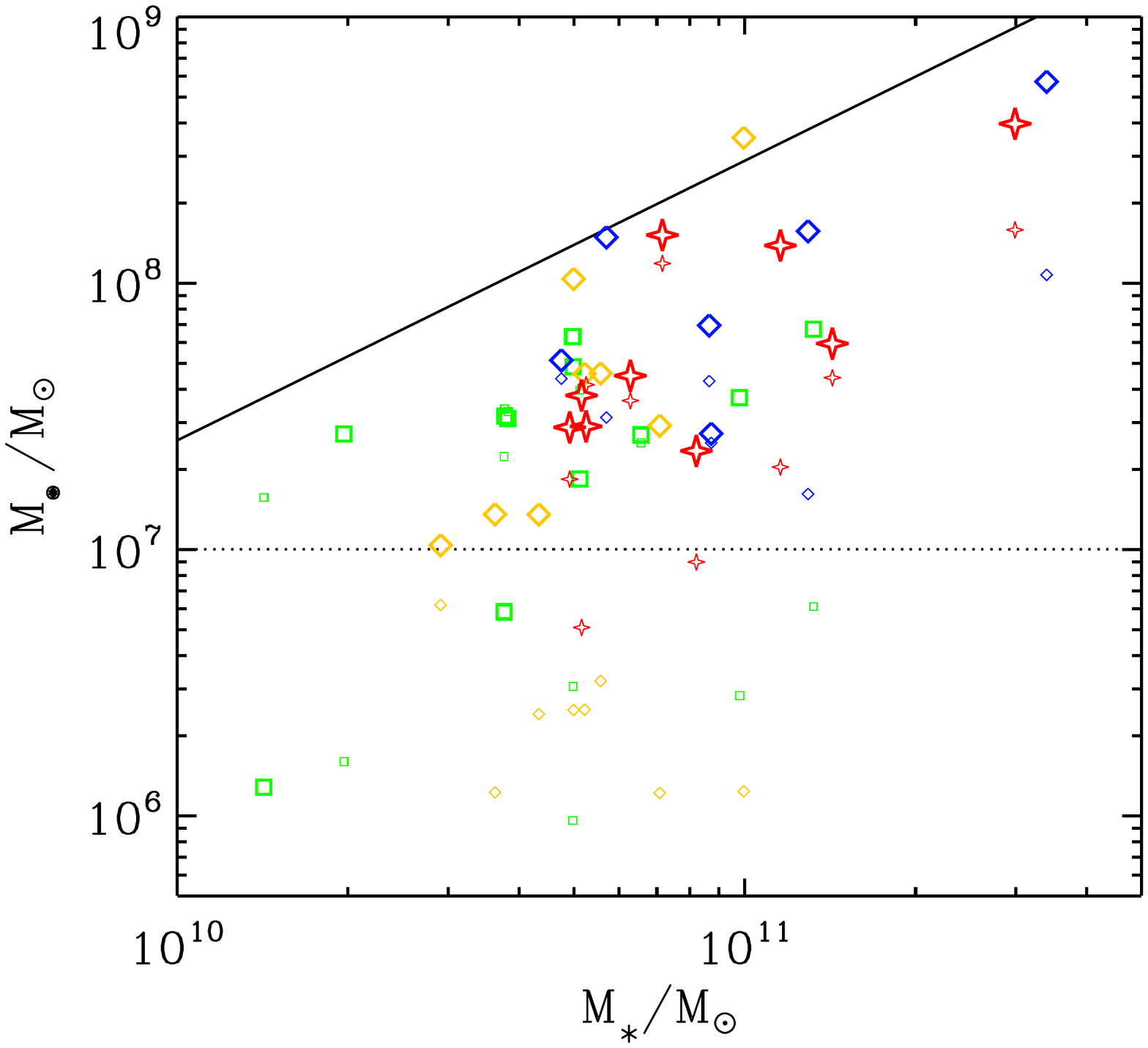}
  \caption{
	Relation between the BH mass and the stellar mass of the host galaxy for our sample of BH pairs.
	The different symbols illustrate the different classes of BH pairs,
	where large symbols correspond to the more luminous BHs in a pair and the small symbols to the less luminous ones.
	The black solid line shows the fit from \citet{McConnell} and the black dotted line marks our resolution limit.
          }
\label{mbh_mst}
\end{figure}

In the following figures, we illustrate dual AGN as red stars,
offset AGN and their less luminous counterparts are represented by blue diamonds, unresolved offset AGN and their counterparts are illustrated by yellow diamonds and dual BHs without AGN are shown as green squares.
The more luminous BH in a pair is always represented by a large symbol and the less luminous one by a smaller one.
Fig. \ref{mbh_mst} shows the masses of the BHs in our sample versus the stellar mass of their host galaxies.
We estimate the stellar mass as the total mass of all stars corresponding to the subhalo identified with SUBFIND (\citealt{Dolag09}, \citealt{Springel_subfind}) which contains the BH.
In some cases SUBFIND still identifies two different subhaloes for the two BHs in a pair although the smaller subhalo  is clearly within the larger one.
In these cases we sum up the stellar mass of both subhaloes.

The black solid line illustrates the best fit for the observations of \citet{McConnell} at $z=0$.
Since we found in \citet{Hirschmann} and \citet{Steinborn_2015} that the Magneticum Pathfinder Simulations agree very well with these observations and that this relation does not change significantly between $z=0$ and $z=2$, we can use the black line also at $z=2$ for orientation.
As expected, almost all BHs lie below the observed relation from \cite{McConnell},
since the merger triggered star formation activity starts earlier than the fuelling of the BHs and the actual BH merger.
This is in excellent agreement with the findings from \citet{Kormendy}.

Fig. \ref{mbh_mst} also shows that 
offset AGN are mostly more massive than their inactive counterparts.
Furthermore,
above $M_{\bullet} \sim 7 \cdot 10^7 M_{\odot}$, all BHs are either
dual or offset AGN.  For lower masses, active and non-active BHs cover
the same mass range.  By definition, the inactive counterparts of
unresolved offset AGN are below the resolution limit of $M_{\bullet} =
10^7 M_{\odot}$ (black dotted line).  These BHs would not necessarily
be inactive, if they were properly resolved.  The stellar mass of
their host galaxies is very large compared to the BH mass, probably
due to the merger.  The BH growth is suppressed, which is very
untypical, since the BHs usually grow very fast right after the
seeding because they are seeded with an artificially low BH mass.
This could explain why the unresolved counterparts of offset AGN
(small yellow diamonds) have even lower BH masses than most BHs in
pairs without AGN (green squares).  We suspect that without this
seeding and resolution effect, our simulation might contain more dual
AGN.

\subsection{Triggering versus suppressing accretion}
\label{triggering_vs_suppressing}
\begin{figure}
  \includegraphics[trim = 0mm 59mm 0mm 60mm, clip,width=0.5\textwidth]{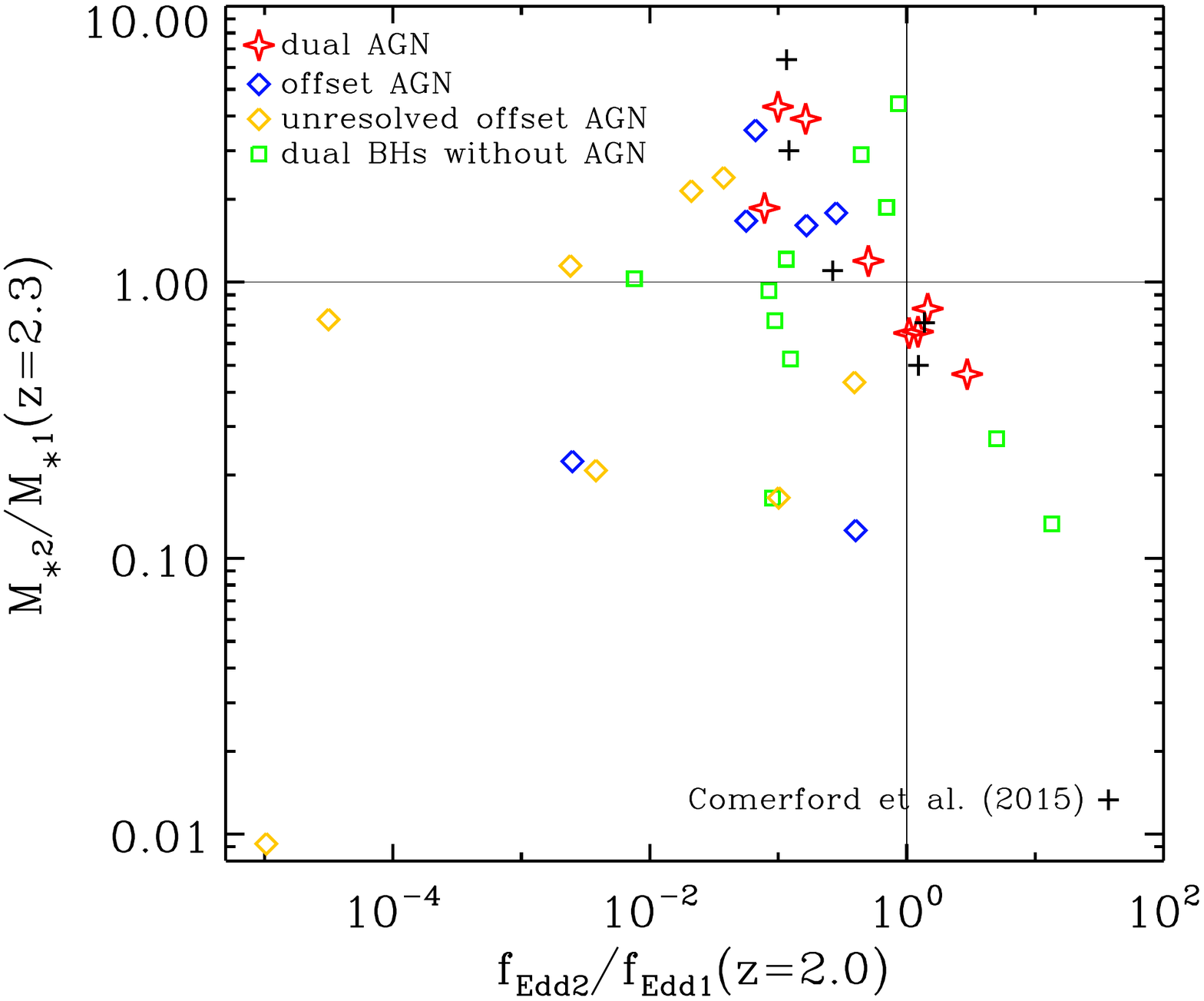}
  \caption{The ratio of the Eddington ratios compared to the stellar mass ratio.
    The indices 1 and 2 correspond to the more and less luminous BH, respectively.
	For all dual AGN the more luminous AGN originates from the less massive progenitor galaxy,
	whereas offset AGN always have a higher Eddington ratio than their less luminous counterpart.
	Our dual and offset AGN lie in the same range as observed ones (black crosses) from \citet{Comerford_2015}.
	Note that we show only 8 dual AGN in this plot because we cannot give a secure estimate for the merger mass ratio of dual AGN number 1 due to a merger of three galaxies.
          }
\label{fedd1_fedd2_mst1_mst2_mgas}
\end{figure}

\cite{Comerford_2015} found that all dual AGN and dual AGN candidates
in their sample have one interesting similarity: The BH in the less
luminous galaxy always has the higher Eddington ratio. 
Since their sample is very small, additional data is needed to constrain this result.
At that point our simulation can be helpful:
In contrast to observations, it has the advantage that we know the exact stellar mass before the merger
and thus we are not biased by effects like stellar stripping.
Although we
study much higher redshifts, we interestingly find the same behaviour
in our simulation for dual AGN. 
To characterize the processes which drive our results we traced our merging systems
0.5 Gyr back in time (to $z=2.3$) to infer the stellar masses of the progenitor 
galaxies.
This is demonstrated in
Fig. \ref{fedd1_fedd2_mst1_mst2_mgas}, showing the stellar mass ratio
of the two progenitor galaxies at $z=2.3$ versus the ratio of the
corresponding Eddington ratios at $z=2$.  The indices 1 and 2
correspond to the more and less luminous BH, respectively.  For dual
AGN, the galaxy with the higher Eddington ratio has the lower stellar
mass ratio and vice versa.  Offset AGN, on the contrary, behave
differently: as expected intuitively, the BH appearing as an AGN
always has the higher Eddington ratio, irrespectively of the host
galaxy mass.  BH pairs without AGN activity can behave like either
dual or offset AGN.  Fig. \ref{fedd1_fedd2_mst1_mst2_mgas} also shows
that the simulation predictions cover the same range as the observed
ones (black crosses).  Together with the findings in Section
\ref{sample}, where we also found similar trends like in observations
at low redshifts, this indicates that the underlying physical
processes which drive the formation of dual and offset AGN are the
same at $z=2$ as at low redshifts.

\begin{figure}
  \includegraphics[trim = 0mm 60mm 0mm 60mm, clip,width=0.5\textwidth]{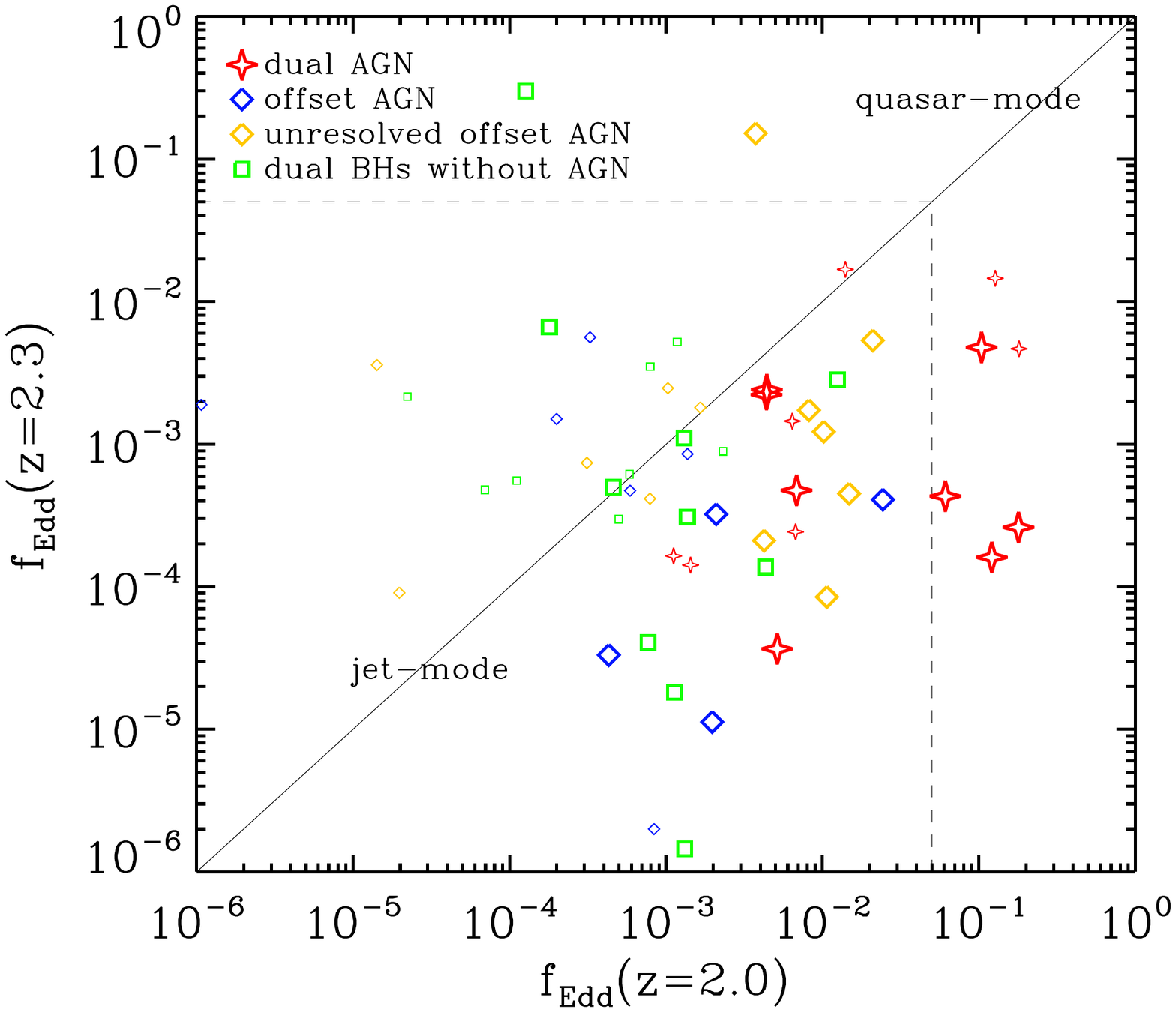}
  \caption{Comparison of the Eddington ratio at $z=2.0$ and $z=2.3$.
	For almost all dual AGN the Eddington ratio increases in this time interval.
	In pairs with only one AGN the Eddington ratio of the AGN increases,
	whereas it decreases for the less luminous BH.
	Dual BHs without AGN scatter in the range up to $f_{Edd} \approx 10^{-2}$. 
	Note that some BH do not yet exist at $z=2.3$ and hence, there are less than 60 data points.
          }
\label{fedd}
\end{figure}

Fig. \ref{fedd} shows a comparison of the Eddington ratios of the progenitor systems at $z=2.3$ and for the BH pairs at $z=2.0$, where the symbols are the same as in Fig. \ref{mbh_mst}.
Non-active BHs scatter over the whole range, but mostly correspond to
the jet-mode (or radio-mode), which is typically defined as
$f_{Edd}<5\cdot 10^{-2}$.  As shown in \citet{Steinborn_2015}, in this
mode the gas reservoir in the vicinity of the inactive BHs is either
heated or consumed, leading to rather low BH accretion rates and thus,
low AGN luminosities.

For dual and offset AGN, the simulations predict one clear difference:
for the majority of dual AGN, the Eddington ratio significantly
increases for both BHs between the progenitors at $z=2.3$ and the pairs at 
$z=2.0$. 
For the offset AGN (both resolved and unresolved) on the contrary,
the Eddington ratio also increases for the active AGN, but
for the non-active counterparts (small blue and yellow diamonds)
it either hardly changes or it even decreases with time (except in one case). This indicates that, 
in contrast to dual AGN, offset AGN might prevent their
counterparts from accreting more gas.

One explanation may be related to the effect of
AGN feedback. Single BHs usually grow rapidly until AGN feedback and
gas cooling are in equilibrium, where feedback leads to lower
accretion rates and gas cooling to higher accretion rates. When the
equilibrium is reached, they grow along the M-$\sigma$ relation
\citep{Churazov}. In BH pairs, feedback from two BHs heats the same
gas reservoir, and hence the heating may dominate compared to gas
cooling, leading to low Eddington ratios. This could prevent the less
massive BH from growing further, as it `starves' because of the
feedback of the more massive BH, whereas the other BH is massive
enough to appear as AGN also at low Eddington ratios.

\begin{figure}
  \includegraphics[trim = 15mm 59mm 0mm 60mm, clip,width=0.47\textwidth]{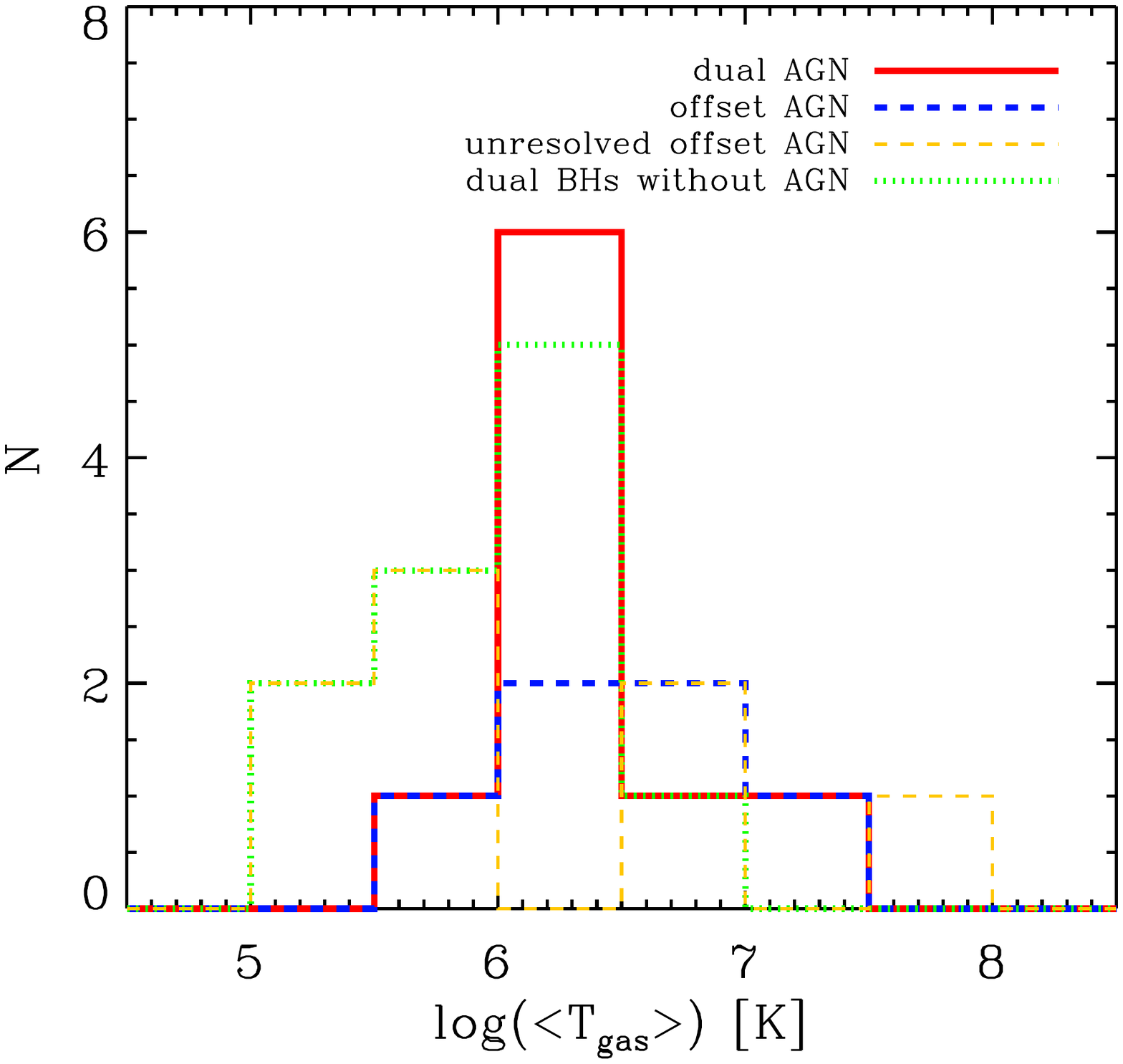}
  \caption{Distribution of the mean gas temperature inside the accretion radius.
  The figure shows only the gas temperatures around the less luminous BH, because we are interested in the mechanisms which suppress AGN activity in the less luminous counterparts of  offset AGN (blue dashed histogram) and BH pairs without AGN (green dotted histogram) in contrast to dual AGN (red solid histogram).
          }
\label{Tmean}
\end{figure}

To understand whether gravity or feedback dominates, we show in
Fig. \ref{Tmean} the distribution of the mean gas temperature inside
the accretion radius\footnote{The accretion radius is set by the most distant gas particle used to calculate the Bondi accretion rate.} of
the less luminous BH.  We are only interested in the less luminous BH,
because whether AGN activity is suppressed in these BHs or not makes
the difference between dual (red solid histogram) and offset AGN (blue
dashed histogram), as we saw in Fig. \ref{fedd1_fedd2_mst1_mst2_mgas}
and Fig. \ref{fedd}.  
We find that most dual AGN have a surrounding
gas temperature between $10^6$K and $10^{6.5}$K. This indicates that
AGN feedback already heats up the surrounding gas, but is still not
strong enough to fully suppress the AGN activity.
There are also two dual AGN surrounded by extremely hot gas ($> 10^{6.5}$K). 
For one of them the stellar masses of the progenitor galaxies are similar such
that none of the BHs can dominate over the other one. In the other
dual AGN pair the more luminous AGN has the less massive progenitor
galaxy.\footnote{At that point we would like to refer the reader to
Appendix A, in which we show that the differences between dual and
offset AGN described in this section are not driven by an overlap of
the accretion radii.}
There are three resolved and three unresolved offset AGN which are
also surrounded by such extremely hot gas, which indicates that
in these cases, AGN feedback is more important in suppressing the
accretion.
In general however, offset AGN (also the unresolved ones) 
are typically characterized by a wider distribution of the surrounding temperatures.
Five of the unresolved counterparts of offset AGN are surrounded by gas
with a lower temperature and hence, the AGN feedback of the second,
more massive and resolved BH, is not relevant. We assume that gravity
might play the most important role in these cases. 

The green dotted histogram in Fig. \ref{Tmean} shows the temperature
distribution around BH pairs without AGN.  Five of these BHs are
hosted by less massive young galaxies which contain a lot of cold gas
but the BHs are simply too small to show AGN activity, although the
distribution peaks in the same range as that of the dual AGN.

Finally, we would like to caution the reader that the gas temperature
is not only affected by AGN feedback.  However, in our cosmological
simulation it is not possible to directly distinguish between
different heating mechanisms, but since we consider only gas inside
the accretion radius, the contribution of AGN feedback is probably
large, especially for very high gas temperatures.

In addition, we expect that not only the mass difference of the
BHs and their associated feedback might play a role, but also the total mass difference of the two
progenitor galaxies, which we will address in more detail in the next 
Section.

\subsection{Merger mass ratio}
\label{triggering}
\begin{figure}
  \includegraphics[trim = 0mm 60mm 0mm 60mm, clip,width=0.5\textwidth]{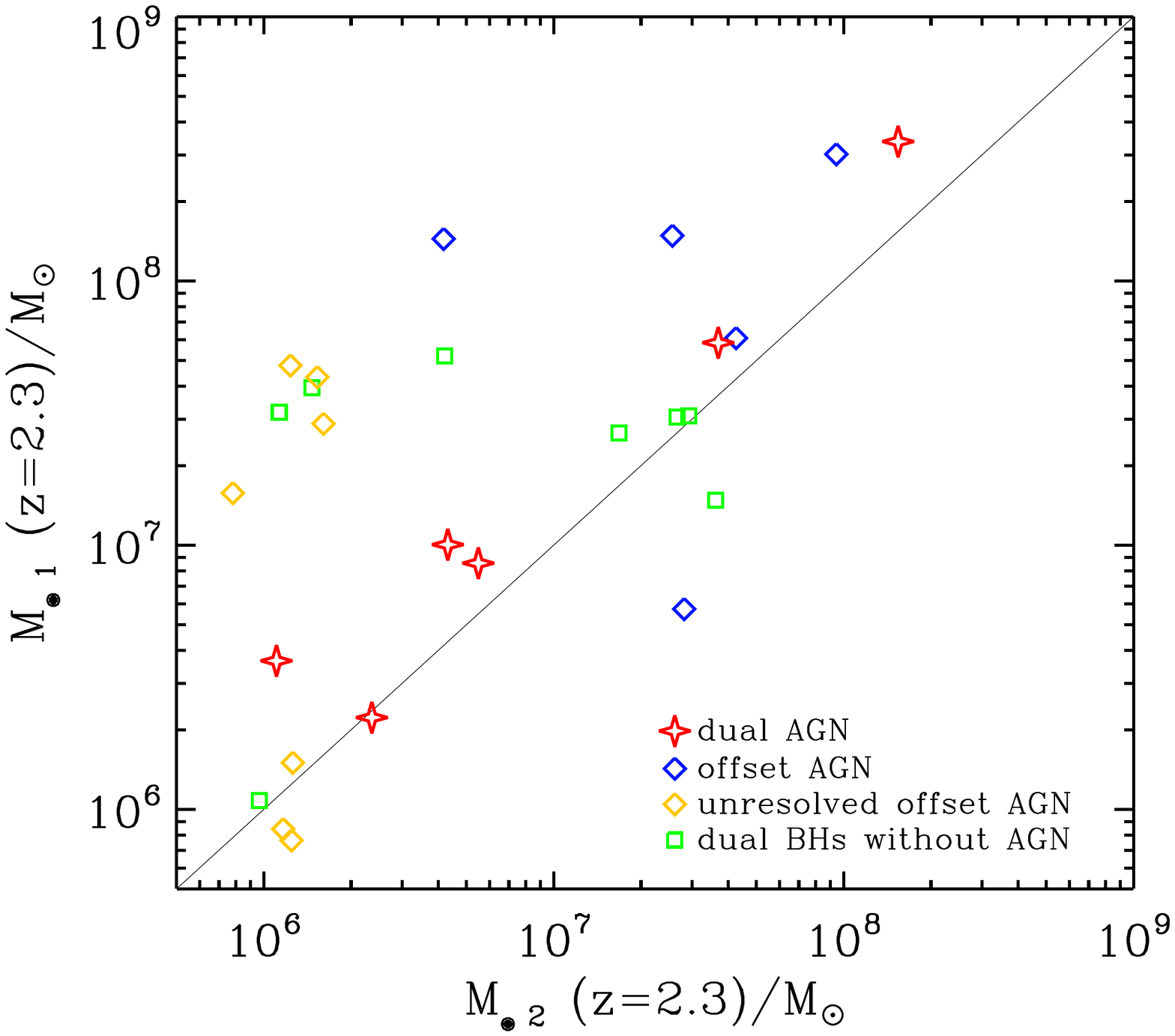}
  \caption{Comparison of the masses of the more ($M_{\bullet 1}$) and the less ($M_{\bullet 2}$) luminous BH around 0.5 Myr before (at $z=2.3$) they were detected as dual BHs.
	For dual AGN the masses are similar, whereas for pairs with one or no AGN they can differ by more than one order of magnitude.
	BHs with masses lower than $5\cdot 10^6 M_{\odot}$ are found in all three types of BH pairs.
	Again, there are less than 60 data points, because some BHs are not yet seeded at $z=2.3$.
          }
\label{mbh}
\end{figure}
To explicitly demonstrate the importance of the mass difference between the two merging BHs for the activity of BH pairs, we compare their masses in the snapshot before the merger, at $z=2.3$ (see Fig. \ref{mbh}).
It is evident that for dual AGN the BH masses are similar, whereas for pairs with one or no AGN, they can differ by more than one order of magnitude.
This indicates that a large difference between the BH masses, and thus also between the galaxy stellar masses, leads to a gas transfer from the smaller to the larger galaxy (e.g. review by \citealt{Barnes_Hernquist_1992}).

That offset AGN are a consequence of mergers with large mass differences, not only in the BH but also in the stellar mass, is supported by Fig. \ref{mhalo_r4th} and Fig. \ref{mhalo_Ngal} in Appendix B, showing environmental dependencies of AGN pairs.
Both dual and offset AGN pairs live in similar, rather high-density regions so that the large-scale environment is apparently not causing the different behaviour of BH pairs. 

\begin{figure}
  \includegraphics[trim = 0mm 67mm 0mm 70mm,
  clip,width=0.5\textwidth]{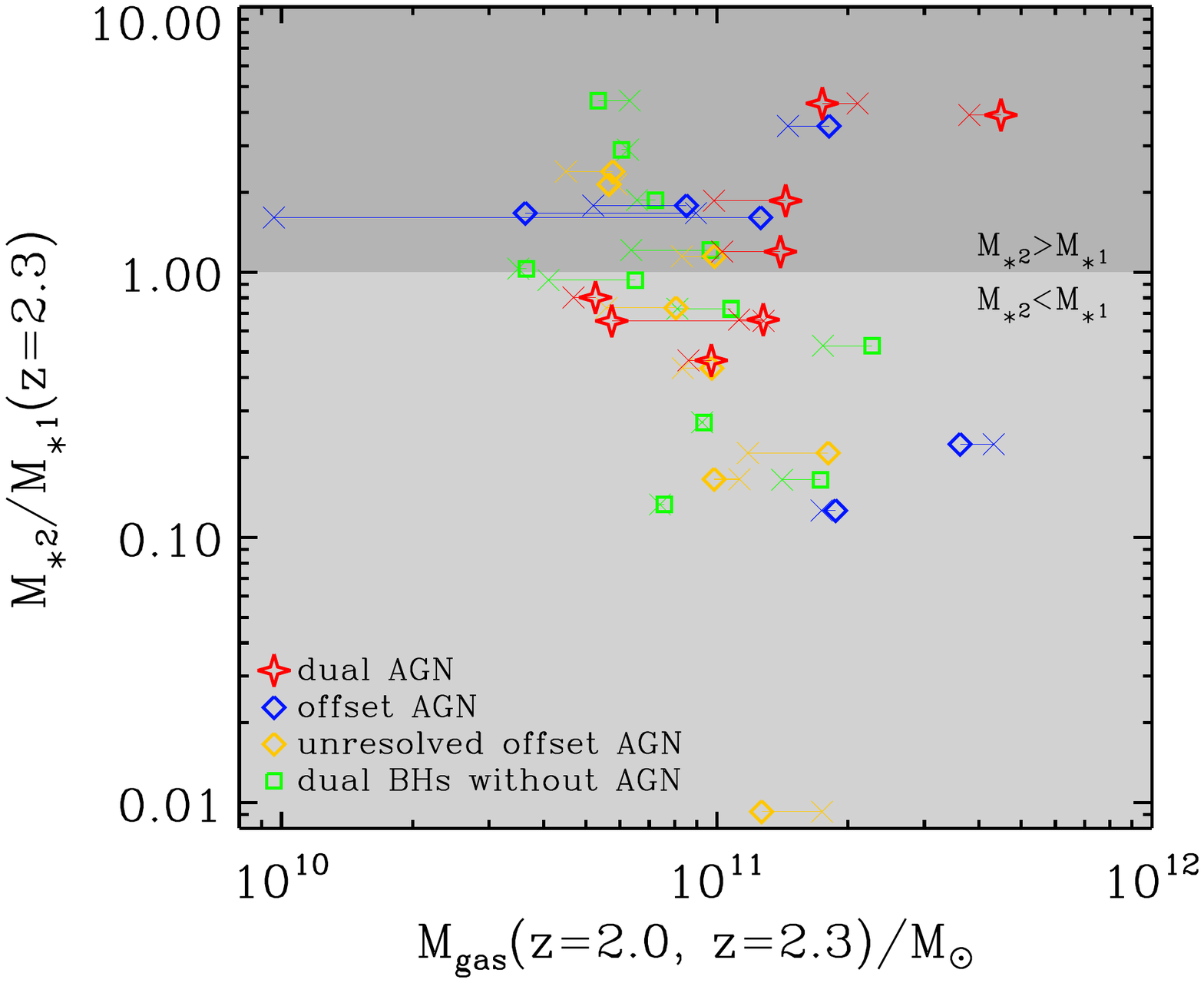}
  \caption{The stellar mass ratio of the progenitor galaxies at $z=2.3$ in comparison with the gas mass of the galaxy hosting the BH pair at $z=2.0$.
  In cases where the two BHs are associated with different subhaloes we sum up their gas masses. The coloured crosses show the gas mass at $z=2.3$, where the lines connect the values at the two different epochs. Like in Fig. \ref{fedd1_fedd2_mst1_mst2_mgas}, we do not plot the dual AGN for which we cannot give a reliable estimate of the merger mass ratio.
          }
\label{mst1_mst2_mgas}
\end{figure}

Fig. \ref{mst1_mst2_mgas} illustrates the merger mass ratios of the progenitor galaxies at $z=2.3$ as a function of the gas masses of the galaxies hosting the BH pairs at $z=2.3$ (coloured crosses) and at $z=2.0$ (other coloured symbols), where the lines connect the values at the two different epochs.
If the two BHs are related to different subhaloes, we sum up the gas masses associated with their galaxies.
Like in Fig. \ref{fedd1_fedd2_mst1_mst2_mgas}, $M_{*1}$ and $M_{*2}$ are the stellar mass of the galaxy with the more and less luminous BH at $z=2.0$.
For values below one, $M_{*2}/M_{*1}$ equals the merger mass ratio, whereas for values larger than one it corresponds to the inverted merger mass ratio and the more luminous AGN originates from the less massive progenitor galaxy.

We find that dual AGN  preferably live in galaxies with a large gas content and a high stellar mass ratio.
There are two exceptions with gas masses lower than $9\cdot10^{10} M_{\odot}$ at $z=2.0$.
In one of these galaxies, the AGN have luminosities of $L_1 = L_2 = 10^{43.2}$ erg s$^{-1}$, just above the threshold for our definition of AGN ($L > 10^{43}$ erg s$^{-1}$).
For the other one, which is represented by the second red cross and the first red star (from left to right) at $M_{*2}/M_{*1} \approx 0.7$, the gas mass decreases by more than a factor of two between $z=2.3$ and $z=2.0$, which is in contrast to the other dual AGN pairs, where the gas mass mostly increases.
Interestingly, this is the only BH pair in our sample which is located in the most massive cluster of the simulation. It has a total halo mass of $M_{\mathrm{halo}}=1.3\cdot 10^{14} M_{\odot}$ and is the only cluster with $M_{\mathrm{halo}}>10^{14} M_{\odot}$ at $z=2.0$.
This might be a hint that AGN are triggered differently in a cluster environment.
We suspect that the gas exchange between galaxies in a cluster might continuously feed the BHs, such that they are not switched off, although the Eddington ratios are relatively low due to the hot environment and the AGN feedback energy is relatively large due to the high luminosity. In the case of our dual AGN pair, the total gas mass of the cluster is $1.4\cdot 10^{13} M_{\odot}$, which is by far larger than the gas mass of the host galaxy only.
Indeed, \citet{Krumpe_2015} find that the mass of the dark matter halo is related to the BH mass.
However, to take a closer look at BH pairs in clusters, we would need 
a larger sample of clusters, which is beyond the scope of this paper.

Fig. \ref{mst1_mst2_mgas} also illustrates that we can roughly distinguish between two different regimes for offset AGN:
A large gas mass would, in principal, allow to feed both BHs, but if one progenitor galaxy is much more massive than the other one, its gravitational potential dominates and hence the BH residing in the more massive galaxy may use up the gas from the smaller one. Thus, offset AGN can have rather low stellar merger mass ratios, in contrast to dual AGN (see also Fig. \ref{mbh}). 
For lower gas masses, the merger mass ratio of dual AGN is still relatively large, i.e., all of them are major mergers, but as $M_{*2}/M_{*1}<1$,
this means that the more massive galaxy contains the more luminous BH.
In contrast, offset AGN are hosted by the less massive galaxy.
For one offset AGN, the gas mass increases by more than one order of magnitude from z=2.3 to z=2. This is not only a consequence of the major merger, but also several minor mergers drive new, cold gas into the center of the host galaxy.

However, there might also be
other mechanisms than those described above which could cause offset AGN, since some offset AGN reside in systems with intermediate gas masses and similar BH and stellar masses.
In these cases, external forces, e.g., related to the environment,
might play a role, which could distribute the gas such that one BH is surrounded by more gas than the other one.

Interestingly, one offset AGN has also a relatively high gas mass, but a high mass ratio -- the typical regime of most dual AGN.
If we take a closer look at this BH pair, we find that the two BHs have a distance of 5.9 kpc with no overlap of the radii used for the calculation of the BH accretion rates.
Furthermore, the more luminous BH (the AGN) was seeded during the last 0.5 Gyr and hence the luminosity peak might be an effect of our seeding model.
Nevertheless, we find an interesting result when considering the progenitor galaxy of the AGN when looking at the galaxy morphology.
As shown by \citet{Teklu_2015}, a good indicator for the galaxy morphology is the so-called $b$-value, which describes the position in the $M_*-j_*$-plane, where $j_*$ is the specific angular momentum of the stars.
The connection between morphology and the position on the $M_*-j_*$-plane was first noticed by \citet{Fall83} and was revisited by \citet{Romanowsky_Fall}, who proposed scaling relations for disks and elliptical galaxies, which would correspond to large and small $b$-values, respectively.
Interestingly, the progenitor galaxy of the AGN has a very low $b$-value of -5.75, which is the lowest one in our sample and hence, the galaxy is very likely a compact spheroidal galaxy.
This comes along with a very low gas mass of $M_\mathrm{gas}=2.1\cdot 10^{10} M_{\odot}$.
The stellar mass of this galaxy is also low with $M_*=7.2\cdot 10^9 M_{\odot}$ and the BH is not yet seeded at $z=2.3$.
The second progenitor galaxy has a larger stellar mass of $M_*=2.6\cdot 10^{10} M_{\odot}$, a larger gas mass of $M_\mathrm{gas}=1.2\cdot 10^{11} M_{\odot}$ and also a higher $b$-value of -4.57.
We suspect that the gas from this galaxy feeds the BH from the smaller spheroidal galaxy during the merger, whereas its own BH already accreted much gas in the past so that the surrounding gas has already been heated up by AGN feedback.

To summarize, the different outliers described in this subsection indicate that there exist no specific conditions for a BH pair to become a dual AGN pair or an offset AGN.
Nonetheless, as expected, both the gas mass and the merger mass ratio are indeed crucial quantities for shaping  different types of BH pairs.

\subsection{Gas accretion history}
\label{trace}
\begin{figure}
  \includegraphics[trim = 21mm 15mm 30mm 20mm, clip,width=0.5\textwidth]{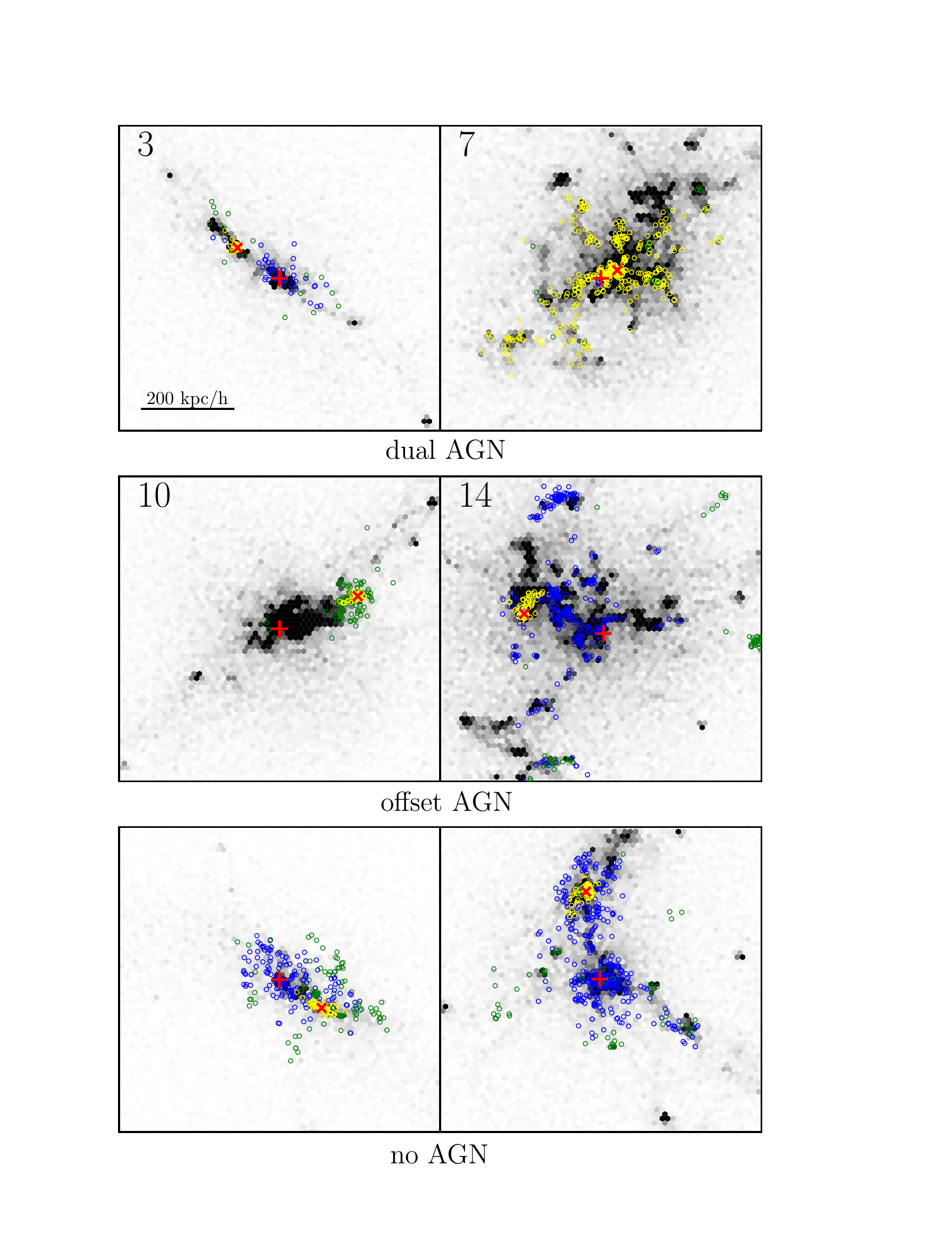}
  \caption{
	The figure shows the progenitor galaxies of two dual AGN pairs, two offset AGN and two BH pairs without AGN at $z=2.3$,
	where the IDs are the same as in Table \ref{dual_AGN_table}, Table \ref{offset_AGN_table} and Fig. \ref{collage}.
	The grey shades visualize the overall gas distribution
	and the coloured circles show where the gas which is accounted for the estimation of the BH accretion rate of the more luminous BH at $z=2$ was located one snapshot earlier, i.e. at $z=2.3$. In that way, we can see where the gas which feeds the more luminous BH comes from.
	Blue and yellow circles indicate gas from the progenitor galaxy of the more and less luminous BH, respectively, whereas green circles represent gas which was associated to none of the two galaxies.
	In every panel, the red plus sign marks the more luminous BH of the pair and the red `x' sign the less luminous one.
          }
\label{tracegas}
\end{figure}

We investigate the evolution of the gas content
feeding the central BHs. Therefore, we trace the gas particles inside
the accretion radius of the more luminous BH at $z=2.0$ back in time to check the origin of these gas particles within the progenitors at $z=2.3$
and exemplarily visualize their spatial distribution 
for two dual AGN pairs, two offset
AGN and two BH pairs without AGN (see Fig. \ref{tracegas}). Each
panel shows a comoving volume of (700kpc/h)$^3$ at $z=2.3$. The red
cross and the red plus sign show the positions of the more and less
luminous BHs (luminosities at $z=2.0$), respectively. The density of
the gas particles is illustrated by the grey shades, while the traced
gas particles are indicated by the coloured circles. In other words,
all of the gas shown as coloured circles will contribute to the
accretion rate of the BH represented as a red plus sign one snapshot
later, i.e. at $z=2.0$.  Blue and yellow circles mark gas which is, at
$z=2.3$, associated to the host galaxy of the more and less luminous BHs
(luminosities at $z=2.0$), respectively\footnote{Note that if the
  blue or yellow gas particles correspond to the main subhalo, it can
  occur that they surround substructures.}, whereas green circles are
gas particles which are associated to none of the two progenitor
galaxies. 

The top left panel shows an example of a dual AGN residing in a gas
filament along which the two galaxies merge.  In this case, gas from
both progenitor galaxies, but also gas from the filament, feeds the
central BHs and thus triggers nuclear activity, although the larger
galaxy contributes more gas than the smaller one. The second example of a dual AGN,
on the top right shows a different case, where gas from only one
progenitor galaxy triggers both AGN.  

For offset AGN we also show two
fundamentally different cases. In the first case, almost all of the gas
that triggers AGN activity originates from the smaller galaxy (see middle left
panel). Alternatively, if the more massive galaxy has a large gas
reservoir, the smaller BH may be simply not massive enough to
``compete" with the massive one. 

The bottom panels show two examples
for BH pairs without AGN. In both cases, gas from both the larger
galaxy and the in-falling substructure as well as gas from outside the
galaxies can contribute to the BH accretion rate.  However, no matter
how large or small the overall gas reservoir around the BH is, the gas
in these inactive BH pairs is typically spread over a larger area,
i.e., the gas density is lower. This, together with the typically low
BH masses, can explain the rather low BH accretion rates, since --
according to the Bondi formalism -- BH accretion rates scale with the
gas density and the BH mass. 

\begin{figure}
  \includegraphics[trim = 15mm 0mm 5mm 0mm, clip,width=0.5\textwidth]{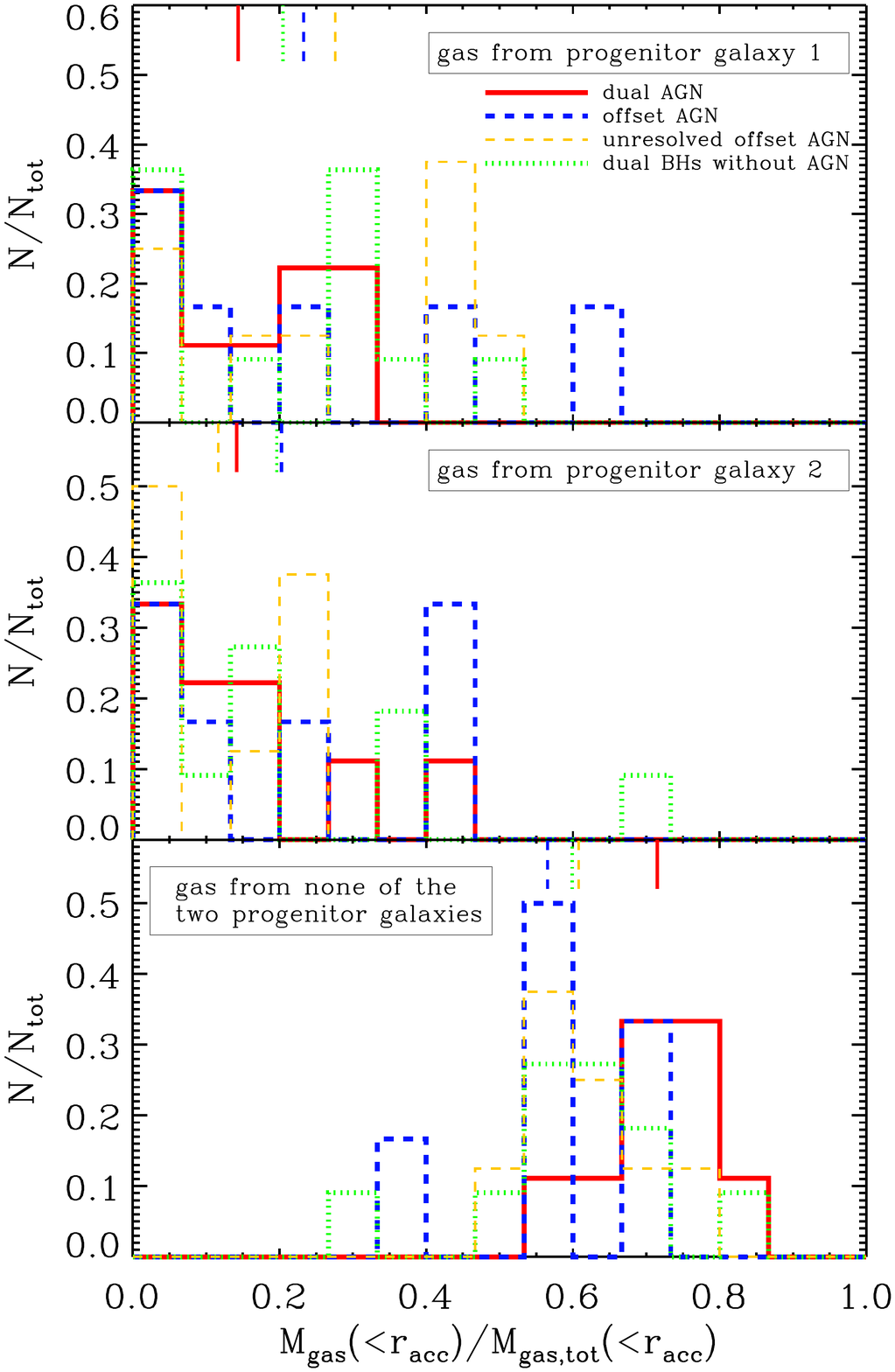}
  \caption{For this figure we traced back all gas particles inside the accretion radius at $z=2.0$ and checked whether they have been in galaxy 1 (progenitor galaxy of the more luminous BH), galaxy 2 (progenitor galaxy of the less luminous BH) or in none of them at $z=2.3$.
	The upper panel shows the gas mass from galaxy 1, the middle panel the gas mass from galaxy 2 and the bottom panel shows the gas mass which was hosted by neither of the two progenitor galaxies with respect to the total mass of the traced gas, respectively.
	The small vertical lines at the top indicate the mean values of each distribution.
          }
\label{trace_hist}
\end{figure}

\begin{figure}
  \includegraphics[trim = 0mm 59mm 0mm 60mm, clip,width=0.5\textwidth]{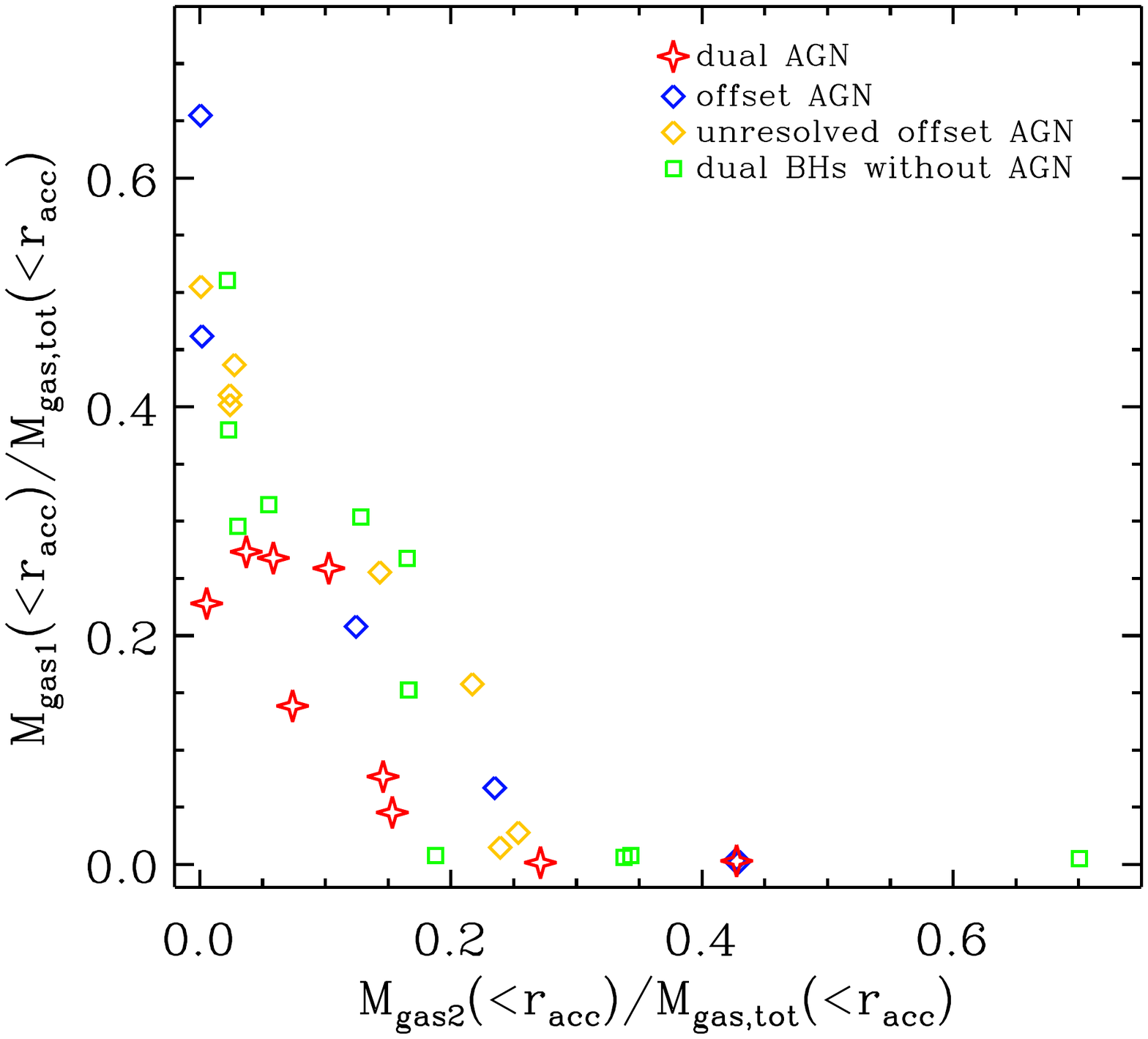}
  \caption{Comparison of the fraction of gas mass with respect to the total gas mass inside the accretion radius at $z=2.0$  which comes from galaxy 1 and galaxy 2.
          }
\label{trace_gal1_gal2}
\end{figure}
To  quantify how strongly dual AGN might be triggered by gas
filaments, we divide the traced gas particles into the following three groups: 
gas which was in progenitor galaxy 1, progenitor
galaxy 2 or in none of them at $z=2.3$.  In the latter case, the gas
was probably accreted by the galaxies between $z=2.3$ and $z=2.0$
either through gas filaments, the infall of gas clumps or additional
minor mergers. Fig. \ref{trace_hist} shows the distribution of
the gas masses belonging to one of these three groups with respect to
the total mass of all gas particles which contribute to the BH
accretion.  We show these distributions for dual AGN (red solid
histogram), offset AGN (blue dashed histogram), unresolved offset AGN
(yellow dashed histogram) and BH pairs without AGN (green dotted
histogram), where we consider only the  more luminous BH in a pair.
The small vertical lines in the top indicate the mean values of each
distribution.  For dual AGN the contribution of gas, which was never
residing in any of the two progenitor galaxies, is clearly enhanced
compared to offset AGN and inactive BH pairs (see lowest panel of
Fig. \ref{trace_hist}).  This indicates that dual AGN indeed accrete
more gas from filaments than the other, less active BH pairs.
Between the other three classes we see no clear differences.
Nonetheless, the contribution of gas from none of the two progenitor galaxies is for almost all BH pairs at least 50\%.

In Fig. \ref{trace_gal1_gal2},
we compare the contribution of the two galaxies to the calculation of the BH accretion with each other.
{Like in Fig. \ref{trace_hist}, it shows that for dual AGN in general less internal gas from the galaxies contributes to the BH accretion than for offset AGN.
In most cases, this internal gas originates from both progenitor galaxies. In contrast,}
for the offset AGN mostly one of the two progenitor galaxies contributes much more gas to the BH accretion than the other one.
This can be the progenitor galaxy of either the less luminous BH or the more luminous one.
In the specific case of unresolved offset AGN, the progenitor galaxy of the more luminous BH is in most cases the main contributor.

\section{Comparison with other theoretical studies}
\label{comparison}
Finally, we would like to put our work in a larger context and compare our results with other theoretical studies, i.e. \citet{Yu_2011}, \citet{vanWassenhove_2012} and \citet{Blecha_2013}, which use either phenomenological models or isolated merger simulations.
However, these methods are different from our ansatz of using cosmological simulations.
\citet{Yu_2011}, for example, construct a phenomenological model for AGN pairs up to $z=1.2$, i.e. they do not make a prediction for $z=2$.
They find that the fraction of dual AGN with respect to the total amount of AGN decreases significantly with increasing redshift up to $z=0.5$.
Above that redshift, the dual AGN fraction seems to saturate, i.e. it does not change significantly in the range $0.5<z<1.2$.
With our simulation we can check whether this saturation holds for higher redshifts up to $z=2$.
The luminosity threshold from \citet{Yu_2011}, above which a BH is defined as an AGN, is
$L_\mathrm{[OIII]}>10^{7.5}L_{\odot}$, which roughly corresponds to $L_\mathrm{bol}>10^{44.6}\mathrm{erg/s}$ assuming $L_\mathrm{bol}/L_\mathrm{[OIII]} \approx 3500$ \citep{Heckman_2004}.
Above this threshold they predict a dual AGN fraction of about 0.02\%-0.06\% for $0.5<z<1.2$.
Our sample contains only one dual AGN above that luminosity threshold corresponding to $\sim 0.05\%$, being in agreement with \citet{Yu_2011}. Hence we indeed find that the saturation holds for larger redshifts. But we caution the reader since this value is based on only one dual AGN pair.

\begin{figure}
  \includegraphics[trim = 10mm 60mm 0mm 65mm, clip,width=0.5\textwidth]{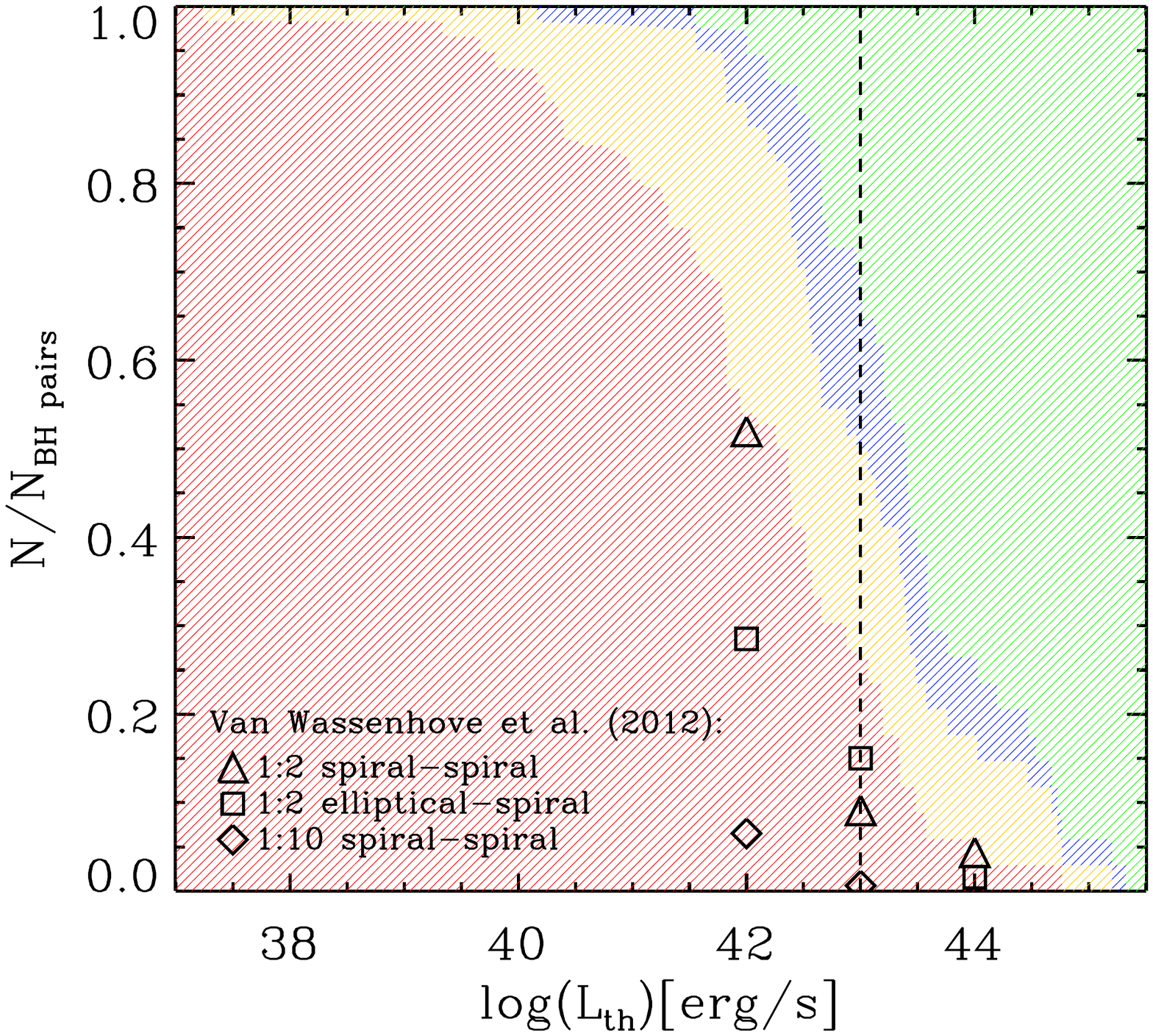}
  \caption{The coloured areas indicate the fraction of simulated dual AGN (red), offset AGN (blue), unresolved offset AGN (yellow) and BH pairs without AGN (green) with respect to the total number of BH pairs, if we choose different luminosity thresholds $L_{\mathrm{th}}$ for our definition of AGN.
  The black dashed line marks the threshold we choose for the analysis in this paper, i.e $10^{43}$erg/s.
  The black symbols show the results from \citet{vanWassenhove_2012} for separations $d>1$kpc with respect to the total amount of dual AGN and offset AGN only.
          }
\label{Lthresh}
\end{figure}
\citet{vanWassenhove_2012} performed three high resolution simulations of isolated galaxy mergers: (i) a 1:2 merger between two spiral galaxies, (ii) a 1:2 merger between an elliptical galaxy and a spiral galaxy and (iii) a 1:10 merger between two spiral galaxies.
The two major mergers are supposed to occur roughly around $z=2$,
while the minor merger occurs around 1 Gyr later.
However, while we can make predictions of the actual number of dual AGN at a given time, \citet{vanWassenhove_2012} can make predictions about the dual AGN fraction only in a temporal manner, i.e. they predict how long a BH pair is a dual AGN. Thus, their predictions always refer only to one specific galaxy merger.

In agreement with our results, \citet{vanWassenhove_2012} find that the gas content and the merger mass ratio play an important role in triggering dual AGN activity. In their simulations, the fraction of dual AGN activity also increases with decreasing separation between the two BHs, which is consistent with our analysis  (see Fig. \ref{d}).

In addition, \citet{vanWassenhove_2012} provide estimates for the dual
AGN fraction with respect to the total time a BH pair spends as either
dual or offset AGN for different luminosity thresholds.  In
Fig. \ref{Lthresh}, we show the contribution of the different types of
our simulated BH pairs with respect to the total number of BH pairs,
depending on the luminosity threshold $L_{\mathrm{th}}$, where the
shaded areas illustrate the contribution of dual AGN (red), offset AGN
(blue), unresolved offset AGN (yellow), and BH pairs without AGN
(green).  Our choice for the analysis in this work, i.e.,
$L_{\mathrm{th}}=10^{43}$erg/s, is marked by the black dashed line.
At that threshold, the contribution of the different types of BH pairs
is similar.  The different black symbols show the results of the three
simulations from \citet{vanWassenhove_2012} for separations larger
than 1kpc (1kpc higher resolution than ours).  Since this is slightly
below our resolution limit, we expect that their dual AGN fraction
might be even lower if they also chose a limit of 2kpc.  Furthermore,
they do not have an upper limit for the BH separations, which should
mainly affect the values at $L_{\mathrm{th}}=10^{42}$erg/s, since they
find that most AGN activity occurs at small separations, in particular
above $L_{\mathrm{th}}=10^{43}$erg/s.  Nevertheless, we see a similar
increase of the dual AGN fraction with decreasing luminosity threshold
like for the 1:2 spiral-spiral merger from \citet{vanWassenhove_2012}.
In contrast to the major mergers, the number of BH pairs in minor
mergers is smaller than our prediction and regarding the
elliptical-spiral merger, the number of BH pairs is also smaller --
except for $L_{\mathrm{th}}=10^{43}$erg/s -- than our estimate. We
suspect that their estimates are in general lower than our
predictions, because their spiral galaxies have a much lower gas
fraction of $f_\mathrm{g}=0.3$ than spiral galaxies in the Magneticum
Simulations at $z=2$, where $f_\mathrm{g} \sim 0.6-0.8$ \citep{Teklu_2015}.  In
addition, the environment in simulations of isolated galaxy mergers is
different to that in cosmological simulations, because there is a
finite gas supply, whereas galaxies in cosmological simulations can
continuously accrete new gas as long as the hot halo does not shut off the gas supply. We conclude that dual AGN activity is
indeed mainly triggered by gas-rich major mergers. Furthermore, we
want to emphasize that we are in very good agreement with the results
from \citet{vanWassenhove_2012} at $L_{\mathrm{th}}=10^{45}$erg/s,
where the effect of the maximum separation should not affect our
results.

Finally, our results agree with those of \citet{Blecha_2013}, who present hydrodynamic simulations of major mergers with different merger mass ratios and gas fractions.
They also find that the dual AGN activity increases with both the merger mass ratio and the gas fraction, although the effect of the mass ratio is not entirely clear, probably due to the fact that they only consider major mergers.
Furthermore, like \citet{vanWassenhove_2012}, they find that the AGN activity is larger in the late phase of a merger, i.e. when the separations between the two BHs are small, which is also in agreement with our results.

\section{Conclusion}
\label{conclusion}
In this study we explore the properties and the origin of dual and
offset AGN, as well as BH pairs without AGN, taking advantage of a cosmological simulation covering
$(182 \mathrm{Mpc})^3$, taken from the set of Magneticum Pathfinder
Simulations. The simulation includes an improved treatment of
super-massive BHs and ran down to redshift $z = 2$. It predicts an
evolution of the AGN luminosity function which is consistent with
observations. The novel treatment of the black holes in the
simulation offers a unique testbed to study the properties and
evolution of BH pairs. At $z = 2$, the simulation contains 34 BH
pairs with a comoving separation smaller than 10kpc. Nine of them are
pairs of dual AGN, 6 are offset AGN and 8 are unresolved offset AGN,
where the mass of the smaller counterpart is not resolved. However,
the remaining 11 BH pairs show no actual AGN activity. 
In the simulation, all BH pairs originate from galaxy mergers independent of whether there is an AGN or not, which implies
that merger activity by itself is not always sufficient to trigger AGN
activity. To investigate the mechanisms which trigger the AGN
activity in detail, we traced the BHs, their progenitor galaxies and
the gas which contributes to the calculation of the accretion rate
back in time.  Our main results are the following:
\begin{itemize}
\item{Almost all BHs in pairs lie below the observed $M_{\bullet}$-$M_*$ relation.
Whilst the most massive BHs appear typically as AGN, less massive ones can be both, either active or non-active.}
\item{We find that the merger mass ratio, the gas mass and the gas accretion history are important factors in triggering dual AGN activity.}
\item{Dual AGN activity dominates for small spatial separations between the two BHs, whereas inactive BH pairs tend to have larger separations.}
\item{In dual AGN pairs, the less massive progenitor galaxy always hosts the BH which later on has the higher Eddington ratio.}
\item{Dual AGN have similar BH masses and grow together, i.e., the Eddington ratio of both BHs in a pair increases during the merger.}
\item{At $z=2.0$, the gas which triggers dual and offset AGN consists mainly of freshly accreted gas, which encounters the progenitor galaxies during or right before the merger, probably through gas filaments or through accretion of smaller substructures.
The contribution of this accretion to the BH accretion is on average clearly larger for dual AGN than for BH pairs with only one or no AGN.}
\item{In most cases, one of the two progenitor galaxies contributes much more gas to the BH accretion than the other galaxy, especially for offset AGN.
This can be either the progenitor galaxy of the more luminous or the less luminous BH.}
\item{Offset AGN can exist in galaxies with relatively low gas masses, if the BH masses are so small that the gas reservoir is still large enough that at least one BH can accrete with a high Eddington ratio, i.e., in the radiatively efficient regime.}
\item{Offset AGN can also be the consequence of a merger of a larger
  with a smaller galaxy, where the large BH accretes and heats up so
  much gas that it either uses up or evaporates the cold gas reservoir
  of the smaller BH. The same effect can be seen in BH pairs without
  AGN, with the difference that there is even 
  too little gas to accrete for the more massive BH for being classified as an AGN.}
\end{itemize}
This study is based on a state-of-the-art large-scale cosmological
simulation for which for the first time not only the resolution is
high enough to resolve galaxies, but also the volume is large enough
to capture the generally rare events of dual AGN. With our sample of
BH pairs we can explain fundamental differences between dual AGN,
offset AGN and inactive BH pairs. However, due to the limited
computational power we can only study the triggering mechanisms down
to $z=2.0$, where galaxy properties like the gas fraction are
significantly different from observed dual and offset AGN at low
redshifts. We expect that this does not qualitatively influence our
results regarding the differences between dual and offset AGN, but
with decreasing redshift quantitative results are very likely to
change. We expect, for example, that the contribution of smooth gas
accretion might be less important for driving nuclear activity at
lower redshifts. Since we find such a process to be essential for
producing dual AGN, we may speculate that in the local Universe a
smaller amount of dual AGN may exist with respect to that of offset
AGN.

In future work we plan to extend this study by investigating AGN trigger mechanisms and the relative role of internal and external processes in global BH populations, and not only for BH pairs. This may shed further light on the still heavily debated question to what extent merger events are responsible (and needed) for driving nuclear activity.

\section*{Acknowledgments}
We would like to thank the anonymous referee for reading the paper very carefully and for giving us very useful comments, which helped us to improve the quality of this paper.
We are especially grateful for the support by M. Petkova through the Computational Center for Particle and Astrophysics (C2PAP).
Computations have been performed at the `Leibniz-Rechenzentrum' with CPU time assigned to the Project `pr83li'.
This research was supported by the DFG Cluster of Excellence `Origin and structure of the universe' and the SFB-Tansregio TR33 `The Dark Universe'.
MH acknowledges financial support from the European Research Council via an Advanced Grant under grant agreement no. 321323‚ NEOGAL.

\bibliography{dual_AGN}

\section*{Appendix A: Intersection between accretion radii}
\begin{figure}
  \includegraphics[trim = 0mm 60mm 0mm 59mm, clip,width=0.5\textwidth]{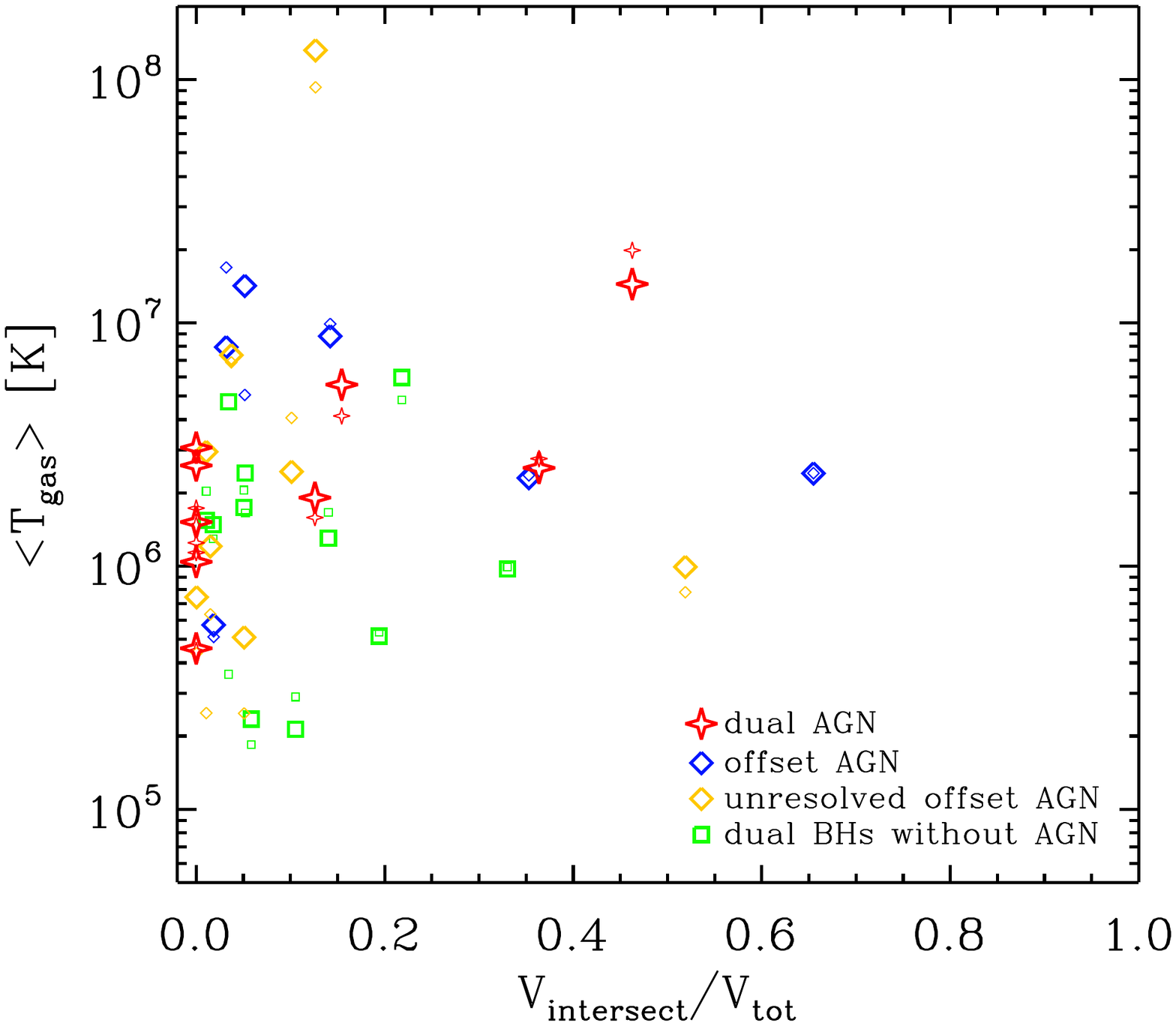}
  \caption{Fraction of the intersection of the volumes used to calculate the accretion rate at $z=2$ with respect to the total volume compared with the mean gas temperature inside the accretion radius of each BH, also at $z=2$.
          }
\label{Vfintersect_Tmean}
\end{figure}

\begin{figure}
  \includegraphics[trim = 0mm 60mm 0mm 60mm, clip,width=0.5\textwidth]{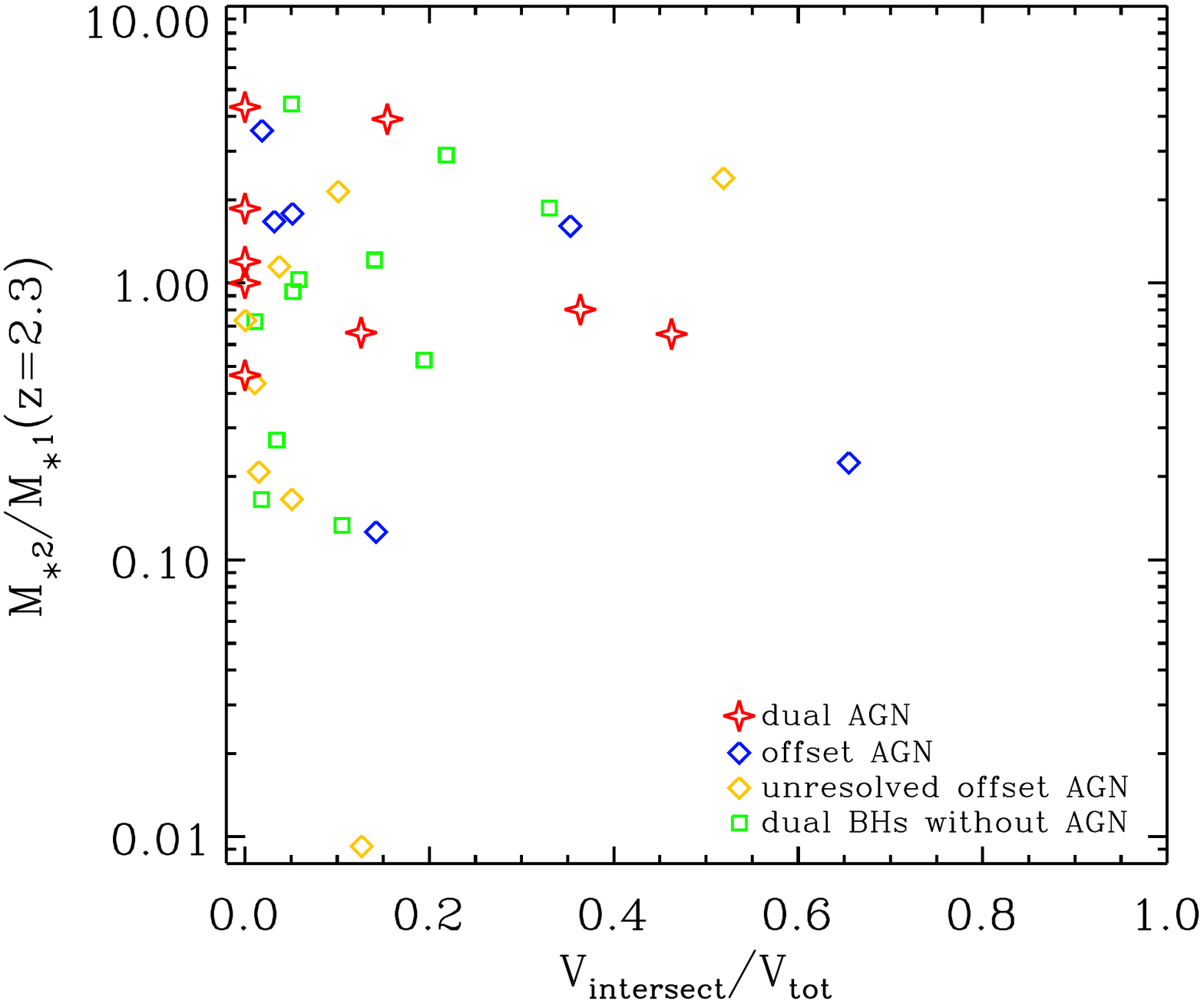}
  \caption{Fraction of the intersection of the volumes used to calculate the accretion rate at $z=2$ with respect to the total volume compared with the stellar mass ratio, where $M_{*1}$ and $M_{*2}$ are the stellar masses at $z=2.3$ and correspond to the more and less luminous BH at $z=2.0$, respectively.
          }
\label{mst1_mst2_Vfintersect}
\end{figure}

Fig. \ref{Vfintersect_Tmean} and \ref{mst1_mst2_Vfintersect} demonstrate that our analysis is not driven by numerical effects due to our BH model.
$V_{\mathrm{intersect}}$ is the volume inside the accretion radius which intersects between the two BHs, and $V_{\mathrm{tot}}=V_1+V_2-V_{\mathrm{intersect}}$ is the total volume.
Since both dual and offset AGN spread over the whole range of $V_{\mathrm{intersect}}/V_{\mathrm{tot}}$ we can rule out that the two different classes are a numerical effect due to our choice of the accretion radius.
There is one trend visible, namely that most dual AGN do not intersect at all.
Many pairs without AGN have a very low gas temperature, because the BH has just been seeded and the galaxy is not very evolved yet. Of course this can be an effect of the seeding model. The few offset AGN with high gas temperatures seem indeed to be triggered by AGN feedback, whereas for the others feedback does not play an important role. The BH masses of the one dual AGN with an extremely high gas temperature are very similar such that none of the two BHs dominates over the other one.

\section*{Appendix B: Environment}
\begin{figure}
  \includegraphics[trim = 0mm 60mm 0mm 60mm, clip,width=0.5\textwidth]{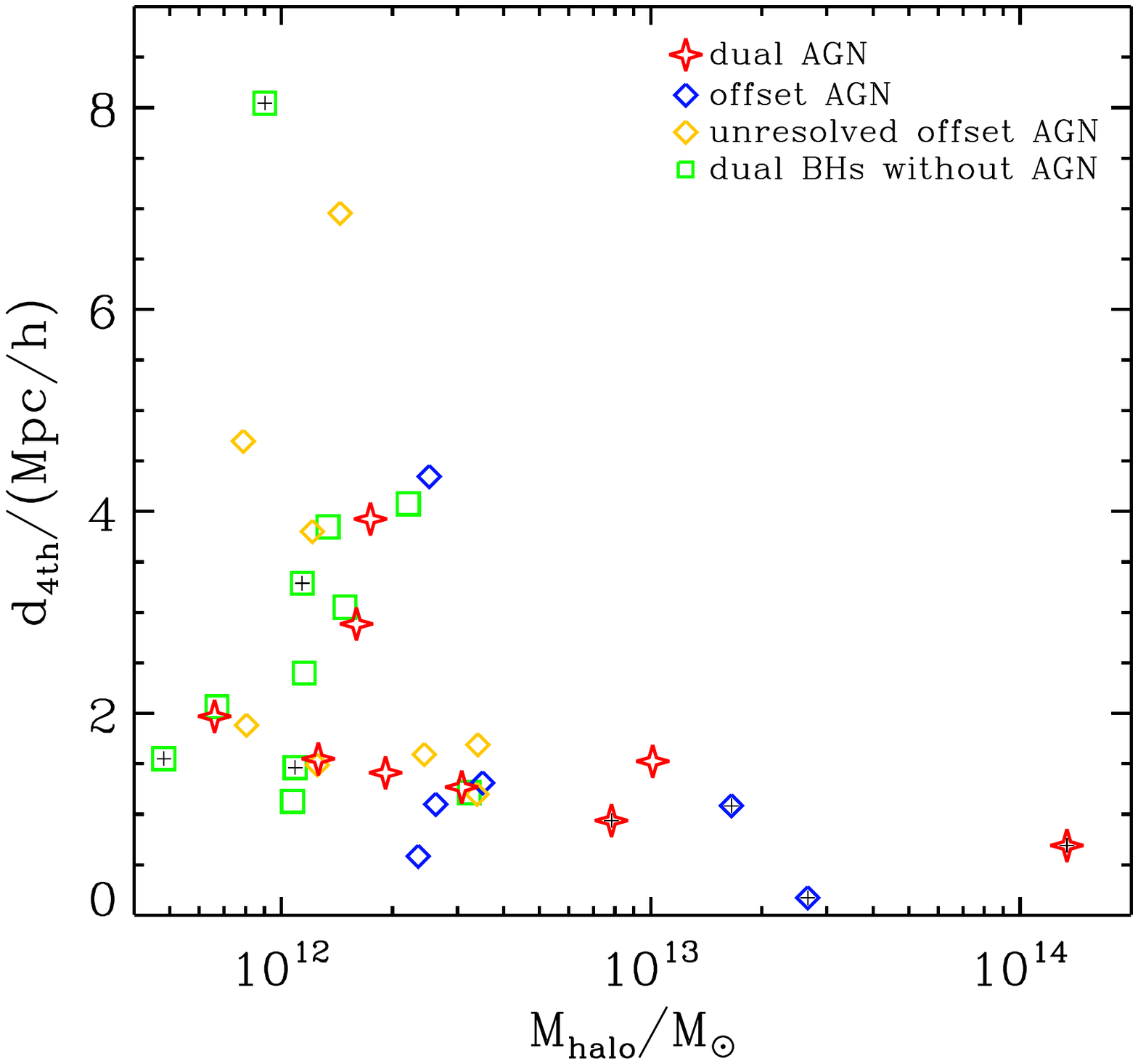}
  \caption{Distance to the forth nearest neighbour galaxy with $M_*>10^{10} M_{\odot}$. The black crosses mark mergers between two substructures, i.e., they do not occur in the central galaxy.
          }
\label{mhalo_r4th}
\end{figure}
\begin{figure}
  \includegraphics[trim = 0mm 60mm 0mm 60mm, clip,width=0.5\textwidth]{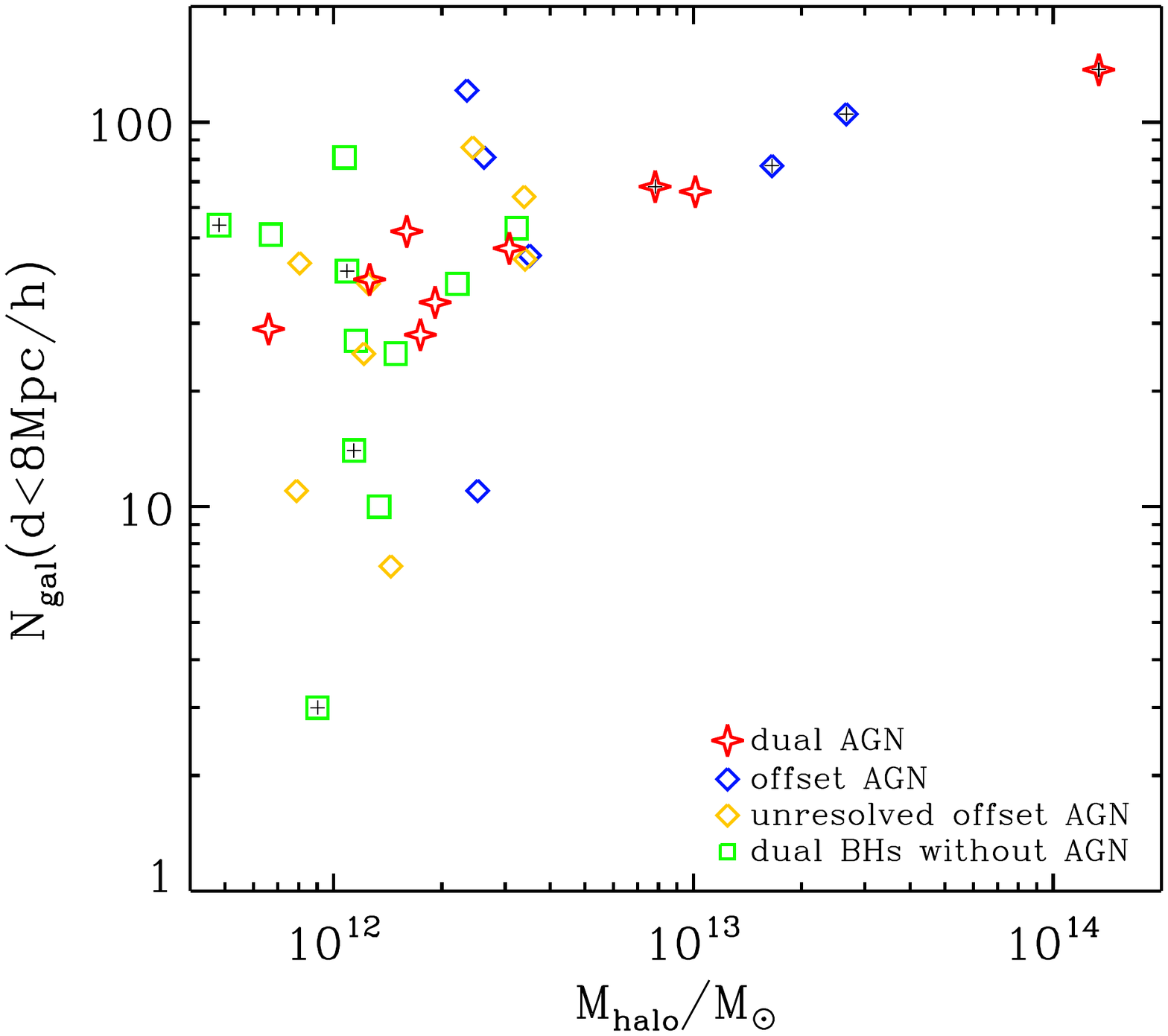}
  \caption{Number of neighbour galaxies within a radius of $8\mathrm{Mpc}/h$ which are larger than $10^{10} M_{\odot}$. The black crosses mark mergers between two substructures, i.e. they do not occur in the central galaxy.
          }
\label{mhalo_Ngal}
\end{figure}
In Fig. \ref{mhalo_r4th} and Fig. \ref{mhalo_Ngal} we show two different measures of the environment against the total the dark matter halo: i) the distance to the 4th nearest galaxy and ii) the number of galaxies within a radius\footnote{We tested several radii and it turned out that using 8 Mpc$/h$ it is best visible that both dual AGN prefer a denser environment.} of $8\mathrm{Mpc}/h$ (e.g. \citealt{Haas_2011}, \citealt{Muldrew_2012} and references therein).
For both approaches we choose a mass threshold of $M_*>10^{10} M_{\odot}$ for the neighbouring galaxies \citep{Baldry_2006}.
Especially in Fig. \ref{mhalo_Ngal} it is visible that both dual and offset AGN prefer a dense environment, although BH pairs without AGN can as well have many neighbouring galaxies.
The two figures also show that all simulated BH pairs in haloes with $M_{\mathrm{halo}} > 4 \cdot 10^{12} M_{\odot}$ contain at least one AGN.
Hence, looking for a dense environment might help to find dual and offset AGN, although we also produce them in less denser environments.

Furthermore, the two figures show clearly that our simulation produces offset AGN only in haloes above $M_{\mathrm{halo}} \approx 2 \cdot 10^{12} M_{\odot}$.
Below that threshold there are only unresolved offset AGN.
In contrast, dual AGN scatter over the whole range of halo masses.
This might be a consequence of our findings that dual AGN often reside in major mergers,
whereas differences in the BH and stellar mass can cause offset AGN.
Thus, in low mass haloes offset AGN need, in contrast to dual AGN, an even less massive counterpart and hence the smaller BH mass is not resolved.
Considering this resolution effect, we do not see a different environment for dual and offset AGN.
However, the approaches used to measure the environment are both spherically symmetric and thus large-scale filaments are not captured by this method.
Hence, it turned out that measuring the gas accretion history like in Fig. \ref{trace_hist} is more useful to investigate AGN triggering mechanisms.

In Fig. \ref{mhalo_r4th} and Fig. \ref{mhalo_Ngal} we also mark mergers between two substructures with black crosses, in contrast to mergers involving the central galaxy.
Although our sample contain only a few groups and only one cluster at $z=2.0$, it is a remarkable result that most of the mergers in these massive haloes occur between two substructures.
This might indicate that, at least at such a high redshift, the central galaxy does not yet play such a dominant role as known from the local Universe.

\bsp

\label{lastpage}

\end{document}